# Remote health diagnosis and monitoring in the time of COVID-19

Joachim A. Behar[1], Chengyu Liu[2,^], Kevin Kotzen[1,^], Kenta Tsutsui[3], Valentina D.A. Corino[4], Janmajay Singh[5], Marco A.F. Pimentel[6], Philip Warrick[7], Sebastian Zaunseder[8], Fernando Andreotti[9], David Sebag[10], Georgy Kopanitsa[11], Patrick E. McSharry[12,13,14], Walter Karlen[15,*], Chandan Karmakar[16,17,*] and Gari D. Clifford[18,19,*]

^equal contribution.

*equal senior authorship.

[1] Faculty of Biomedical Engineering, Technion-IIT, Haifa, Israel

[2] The State Key Laboratory of Bioelectronics, School of Instrument Science and Engineering, Southeast University, Nanjing, China

[3] Department of Cardiovascular Medicine, Saitama International Medical Center, Saitama, Japan

[4] Department of Electronics, Information and Bioengineering, Politecnico di Milano, Italy

[5] Independent researcher, Japan

[6] Department of Engineering Science, University of Oxford, Oxford, UK

[7] PeriGen, Cary, NC, and Westmount, QC, Canada

[8] Faculty of Information Technology, FH Dortmund, Dortmund, Germany

[9] Sensyne Health, Oxford, UK

[10] Independent physician, Grenoble, France

[11] ITMO University, Saint Petersburg, Russia

[12] Carnegie Mellon University Africa, Kigali, Rwanda

[13] African Centre of Excellence in Data Science, University of Rwanda, Kigali, Rwanda

[14] Oxford Man Institute of Quantitative Finance, Oxford University, Oxford, UK

[15] Mobile Health Systems Lab, Department of Health Sciences and Technology, ETH Zurich, Zurich, Switzerland

[16] School of Information Technology, Deakin University, Geelong, Australia

[17] Department of Electrical and Electronic Engineering, University of Melbourne, Melbourne, Australia

[18] Department of Biomedical Informatics, Emory University, Atlanta, GA, USA

[19] Department of Biomedical Engineering, Georgia Institute of Technology and Emory University, Atlanta, GA, USA

**Abstract**

Coronavirus disease (COVID-19), caused by the severe acute respiratory syndrome coronavirus 2 (SARS-CoV-2), is rapidly spreading across the globe. The clinical spectrum of SARS-CoV-2 pneumonia requires early detection and monitoring, within a clinical environment for critical cases and remotely for mild cases, with a large spectrum of symptoms. The fear of contamination in clinical environments has led to a dramatic reduction in on-site referrals for routine care. There has also been a perceived need to continuously monitor non-severe COVID-19 patients, either from their quarantine site at home, or dedicated quarantine locations (e.g., hotels). In particular, facilitating contact tracing with proximity and location tracing apps was adopted in many countries very rapidly. Thus, the pandemic has driven incentives to innovate and enhance or create new routes for providing healthcare services at distance. In particular, this has created a dramatic impetus to find innovative ways to remotely and effectively monitor patient health status. In this paper, we present a review of remote health monitoring initiatives taken in 20 states during the time of the pandemic. We emphasize in the discussion particular aspects that are common ground for the reviewed states, in particular the future impact of the pandemic on remote health monitoring and consideration on data privacy.

**Introduction**

Coronavirus disease (COVID-19) is caused by severe acute respiratory syndrome coronavirus 2 (SARS-CoV-2) that has rapidly spread across the globe. It is a newly recognized illness that initially surged exponentially in Wuhan, China, and from there, to other provinces in China and then across the globe (Yang *et al* 2020, Wang *et al* 2020b, 2020a, Zhou *et al* 2020). As of 28 June 2020, the number of COVID-19 patients around the world surpassed 10 million, with over 500,000 casualties. The clinical spectrum of SARS-CoV-2 pneumonia ranges from mild to critically ill cases and requires early detection and monitoring, within a clinical environment for critical cases and remotely for mild cases. The fear of contamination in clinical environments has led to a dramatic reduction in on-site referrals for routine care. There has also been a perceived need to continuously monitor non-severe COVID-19 patients, either from their quarantine site at home, or a dedicated quarantine location (e.g., hotel). There has also been a need to find alternatives routes to provide regular healthcare services to non-COVID-19 patients while in quarantine. This has created a dramatic impetus to find innovative ways to remotely and effectively monitor patient health status.

Remote health diagnosis and monitoring relates to the diagnosis and monitoring of individuals outside the classical hospital environment, typically in their home but also through mobile clinics. It enables monitoring of patient well-being (e.g., sleep quality), diagnosis of medical conditions (e.g., atrial fibrillation) as well as to track over time changes requiring medical attention. In the context of this review we consider both remote single or repeated measurements for the purpose of diagnosis and/or monitoring. The diagnosis and monitoring actors that were considered were medical professionals or a government body (e.g. ministry of health) or individuals self-monitoring themselves. With the advent of COVID-19, researchers, entrepreneurs, governments and industries around the world have been engaged in inventing or adapting existing technologies to support healthcare in dealing with the novel coronavirus. In many ways, the pandemic has underscored the importance of remote health monitoring in our modern, connected and digital societies. The current work provides an overview of the innovations that have been springing up globally within the context of the COVID-19 pandemic.

Experts from different regions were invited to contribute to this review paper and asked to contribute their opinion on the role, usage and development of remote health monitoring during the time of the pandemic. They categorized the initiatives that developed technologies which: (1) facilitate contact tracing and the reconstruction of exposure to potentially infected COVID-19 patients, with the two subcategories of proximity and location tracing; (2) perform telemonitoring, i.e. use telemedicine tools like video chat and connected medical sensors to diagnose and manage patients; (3) were used in the creation of mobile clinics, where medical services can be accessed away from regular contact points, such as hospitals; and (4) enable point-of-contact screening to rapidly measure symptoms to identify potential patients in everyday locations, such as airports. Furthermore, the experts were asked to review and distinguish between initiatives from the government, industry and academia and what privacy considerations went into their implementation. While performing their review, the authors were asked to focus on the singularity of approaches/initiatives taken in their respective countries, highlight the challenges faced and reason for the lack of uptake of such systems in certain cases. Finally, each author was asked to provide their perspective on the consequences of the pandemic on the future of remote health monitoring in these countries. Although this article is written in the context of COVID-19 the scope is not specific to remote health monitoring for COVID-19 but to the usage of remote health monitoring during the time of COVID-19 i.e. including patient care for chronic or any other conditions while they are in quarantine. This is a non-systematic review which aims to highlight specific initiatives taken in 20 different countries (Figure 1). Whenever possible we attempted to cite peer-reviewed publications covering the initiatives that were highlighted. However, because of the timely nature of this topic, not all such initiatives are yet published in scientific journals or as technical reports.

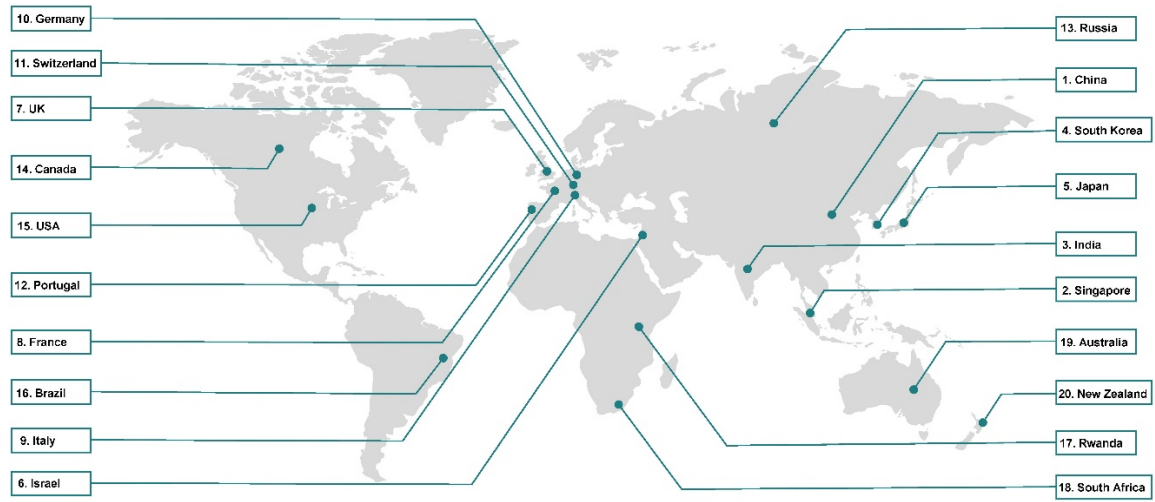

**Figure 1:** Map showing the 20 states and their accompanying number in the review.

**Review of remote health initiatives**

*1. China:* China, the first country to encounter the disease, was also the first to develop innovative remote monitoring methods to assist clinical staff and public health experts. Perhaps the most obvious remote monitoring technology most citizens possess, is the smartphone, allowing everything from telemedicine consults and symptom checking from apps and websites to approximate heart rate monitoring through the camera, and location tracking to identify disease hotspots or contact-tracing. The Chinese government teamed up with Tencent and Alibaba to develop color-coded health rating systems, which are added to the existing apps of WeChat developed by Tencent (Shenzhen, China) and AliPay developed by Alibaba Group Holding Limited (Hangzhou, China). Citizens in hundreds of cities have been asked to download the software that transmits their location to several authorities, including the local police. The app combines geotracking with meta-data, such as travel bookings, to designate citizens with color codes ranging from green (low risk) to red (high risk). High-risk individuals may be banned from certain locations, such as apartment complexes, grocery stores and their own places of employment. Some provinces have different presentations for the color-coded health rating systems, such as the "Sukang" code used in the Jiangsu Province, to facilitate the local government's management in the early stage. Aiming to help the provinces recognize each other's health codes and facilitate travel, as well as to help the flow and sharing of health information, addressing the issue of personal privacy protection, on 29 April 2020, the Standardization Administration of China and the State Administration for Market Regulation jointly released a series of National Guidelines for Personal Health Information Codes (PHI-Code Standards) that specify requirements for health code apps in their collection, processing, and use of personal health information during the COVID-19 pandemic[1]. The PHI-Code Standards include standards on a reference model (GB/T 38961-2020)[2], the data format (GB/T 38962-2020)[3] and an application interface (GB/T 38963-2020)[4] for a unified PHI-Code, which ensure the collection, processing, and utilization of personal health information complies with the national standard of Information Security Technology Personal Information Security Specification (GB / T 35273-2020)[5]. Four types of data will be collected for the PHI-Code system: (1) personal identification information (including name, gender, nationality, identity card number, household registration, address, contact number, medical history, etc.); (2) personal health information (including body temperature, symptoms, period of stay in high-risk areas, contact with high-risk individuals, etc.); (3) travel history; and (4) health assessment information (including medical test results, assessment results, risk level, etc.). According to the new guidelines, health codes must be encrypted and stored using an algorithm satisfying the requirements for national password management. Personal health information services and apps must obtain the express consent or authorized consent of users when collecting data and must keep private content confidential. Both WeChat and AliPay are also being used to track people's movements and ascertain whether they have been in contact with an infected person, in which case, individuals are ordered to self-isolate themselves. China has launched an online consultation system in commonly used mobile apps, such as WeChat and Alipay, to provide technical support for remote screening of COVID-19 infection risk. In addition, Universities and research institutions also contributed to the online consultation systems. An example is that, the second affiliated Hospital of Xi'an Jiaotong University which has developed a free online health consultation and COVID-19 risk screening platform. Besides the inspection of travel history, temperature screening checkpoints have been established at locations with high-density passenger flows, such as train stations, bus stations, railway stations, airports, supermarkets and residential area entrances, which is a necessary, easily implemented, and rapid method of screening large numbers of the population for possible COVID-19 infections (Peng *et al* 2020). Anyone entering a public area is temperature-screened using non-contact infrared thermometers for rapid body temperature measurement. Since core temperature is particularly hard to assess without an invasive probe, far-infrared (FIR) thermal imaging is a key focus for remote/non-contact monitoring in this context. Guidelines require temperatures within normal range i.e., below 37.3 °C (Liu *et al* 2020). Persons whose body temperature exceeds 37.3 °C are prohibited from entering the public ranges and report to the local health management department according

---

[1] Zhang L 2020 China: Recommended National Guidelines for Health Code Apps Issued Libr. Congr. Online: https://www.loc.gov/law/foreign-news/article/china-recommended-national-guidelines-for-health-code-apps-issued/ Retrieved: 08 July 2020

[2] Standardization Administration of China and the State Administration for Market Regulation 2020 Personal health information code-Reference model (GB/T 38961-2020) Code China Online: https://www.codeofchina.com/standard/GBT38961-2020.html Retrieved: 08 July 2020

[3] Standardization Administration of China and the State Administration for Market Regulation 2020 Personal health information code-Data format (GB/T 38962-2020) Code China Online: https://codeofchina.com/standard/GBT38962-2020.html Retrieved: 08 July 2020

[4] Standardization Administration of China and the State Administration for Market Regulation 2020 Personal health information code-Application interface (GB/T 38963-2020) Code China Online: https://www.codeofchina.com/standard/GBT38963-2020.html Retrieved: 08 July 2020

[5] Standardization Administration of China and the State Administration for Market Regulation 2020 Information security technology - Personal information security specification Online: https://www.chinesestandard.net/PDF/English.aspx/GBT35273-2020 Retrieved: 08 July 2020

to the procedures (Cai *et al* 2020). In contrast with the traditional tools such as temperature guns, Guide Infrared (Wuhan, China) supplied automatic fever screening systems, which estimate the temperature of passengers at a distance of several meters, avoiding the close contact required by temperature guns and reducing the risk of infection. Powered by the AI-enabled facial detection technology, the system can automatically focus on a passenger's face and give an alarm when a person with a fever is identified, giving inspectors an ideal tool to deal with densely populated and fast-moving scenarios. Hanwang Technology, also known as "Hanvon" (Beijing, China) supplied a facial recognition system that can identify people even when they are wearing a protective mask with a recognition rate of 95% [6]. Baidu (Beijing, China) has also developed a system that can examine up to 200 people per minute and detect out-of-range body temperatures, without disrupting passenger flow. This system is currently in use at Beijing's Qinghe Railway Station. Robots and smart devices help monitor patient vital signs remotely without person-to-person contact. In the Hongshan Sports Center in Wuhan, the Cloud Ginger (also known as XR-1) robot and the Smart Transportation Robot, which can carry food and medicine from healthcare providers to patients, were specifically modified to assist in the COVID-19 fight. Robots are also being used to diagnose fever in many hospitals. Patients, doctors and nurses wear smart bracelets that are synced with CloudMinds' platform (Shenzhen, China), to monitor their vital signs, including temperature, heart rate and blood oxygen levels, with the aim of catching any early signs of infection. The system has not been validated yet, and it is worth noting that wrist-worn wearables are notoriously noisy and poor at measuring these types of vital signs. Perhaps more notably, a machine developed by engineers at Tsinghua University in Beijing, consists of a robotic arm on wheels that can perform ultrasounds and mouth swab tests and listen to lung sounds[7]. Lung sounds (such as "crackles and rales") can be indicators of fluid buildup in the lungs and be a reason for urgent admission to critical care facilities or administration of steroids or other medications.

Computed tomography (CT) plays a crucial role in the diagnosis and evaluation of COVID-19. However, to avoid potential cross-infection and to facilitate the imaging process, portable CT devices are needed in the bed-wards, and all CT images must be uploaded into picture archiving and communications systems (PACS) for immediate reporting. Mobile CT devices from the Shanghai-based Siemens Healthineers were transported to Huanggang City, Hubei Province, the frontline of the fight against the COVID-19 outbreak. Alibaba (Hangzhou, China) has developed a COVID-19 diagnosis tool that can analyze CT images and provide a diagnosis in the inpatient wards. The facility adopted the multiple thermal insulation structure and used a lead composite layer to protect against X-ray radiation, making it flexible and easy to transport. 5G technology was also applied to ensure the high-speed transmission of medical images for remote diagnosis. For example, the radiology department of the West China Hospital of Sichuan University used 5G for remote CT scanning of COVID-19 patients, which provided strong support for subsequent diagnosis and treatment. Similarly, ultrasonic equipment and remote diagnosis service also obtained the support of a 4G/5G technique. Changzheng Hospital in Shanghai has applied the ultrasonic remote consultation with the support of a 4G/5G network. Medical experts from the remote ultrasound medical center of Zhejiang Provincial People's Hospital used 5G technology to remotely control the ultrasonic robot in the module hospital of Wuhan Huangbo gymnasium to conduct ultrasonic examinations. In these situations, ultrasound experts can complete real-time scanning through remote operation of ultrasonic robot manipulators, thereby effectively relieving the shortage of ultrasound doctors in isolation wards and reducing infection risks (Jia LT, Zhao JQ, Zhang SQ 2020).

---

[6] Clarke L 2020 China's Hanwang Technology claims it can ID mask-wearing faces NS TECH Online: https://tech.newstatesman.com/security/hanwang-technology-id-mask-wearing-faces Retrieved: 30 June 2020
[7] E&T 2020 Robotic arm could save lives on medical frontline amid coronavirus outbreak E&T Online: https://eandt.theiet.org/content/articles/2020/03/robotic-arm-could-save-lives-on-medical-frontline-amid-coronavirus-outbreak/ Retrieved: 30 April 2020

*2. Singapore:* Telemedicine offers greater convenience and improved accessibility to medical support and medication through new digital options. Over the last decade, Singapore has seen a growth in the telemedicine domain, with public and private players emerging to offer a variety of services to patients. Mobile medicine provides greater accessibility for patients who, for a variety of reasons, are unable to attend a hospital or a clinic, by bringing medical practitioners to their bedside. During COVID-19, mobile medicine services are quickly expanding. Singapore took relatively rapid action to restrict the spread of COVID-19. Firstly, to minimize the risk of hospital-based transmission of COVID-19, the Ministry of Health of Singapore has restricted the movement of patients and staff across hospitals[8]. Strict social distancing measures have also been put in place, including advising the public against large social gatherings to mitigate the risk of community transmission[9]. Meanwhile, the country's Integrated Health Information System agency[10] collaborated with KroniKare, to develop iThermo systems, an Android application that connects with a thermal and laser camera, for temperature and distance measurement. The system is designed to identify fever for people in motion and at a distance. Compared to hand-held infrared temperature devices, this system can significantly reduce the risk of infection due to exposure for the user. There are vague claims in the media around the use of artificial intelligence (or more accurately machine learning and computer vision techniques), but beyond this hardware overview, no details on the system's technology performance are available. Perhaps most notably, the country's TraceTogether (Cho *et al* 2020), developed in collaboration with the Government Technology Agency and the Ministry of Health (MOH), provided the first Bluetooth solution in the country to facilitate tracing of close contact with an infected user[11]. The proximity tracing app aims to use Bluetooth signals to determine when two phones are within two meters, or within five meters for 30 minutes. Both phones exchange anonymized temporary IDs (generated by encrypting the User ID with a private key that is held by MOH) together with information about the nearby phone's model, Bluetooth signal strength, and time. The temporary ID can only be decrypted by MOH, with MOH's privately held key. The temporary ID is renewed frequently, making it difficult for anyone (other than the MOH) to identify or link the temporary IDs to any given user. All this information is stored locally on the user's phone, and not sent to MOH. If MOH needs to share a user's proximity data for contact tracing, they seek the user's

---

[8] Lim J 2020 Coronavirus: Doctors, staff and patients to restrict movements to within one hospital The Straits Times Online: https://www.straitstimes.com/singapore/health/coronavirus-doctors-staff-and-patients-to-restrict-movements-to-within-one-hospital Retrieved: 08 July 2020

[9] Sim D 2020 Coronavirus: why did Singapore have more cases than Hong Kong – until now? South China Morning Post Online: https://www.scmp.com/week-asia/health-environment/article/3050039/coronavirus-why-did-singapore-have-more-cases-hong Retrieved: 08 July 2020

[10] GCT 2020 Innovations in Singapore's COVID-19 Response. Glob. Cent. Technol. Innov. Sustain. Dev. Online: https://sgtechcentre.undp.org/content/sgtechcentre/en/home/blogs/covid19response.html Retrieved: 08 July 2020

[11] Hui M 2020 Singapore wants all its citizens to download contact tracing apps to fight the coronavirus Online: https://qz.com/1842200/singapore-wants-everyone-to-download-covid-19-contact-tracing-apps/ Retrieved: 01 May 2020

consent by directly calling the user and providing the user with a code to match with a corresponding verification code on the user's TraceTogether app. The app does not track an individual's location or contacts and data is stored locally on the phone for only 21 days and is not accessed unless the user is identified as having been close enough to another user of the app that was identified as COVID-19 positive at some (unspecified) window around the time of proximity. There is no way for users to self-report tests, and it is not clear how the MOH identifies who is COVID-19 positive, or how accurate the test must be, in order to initiate contact tracing.

*3. India:* India had relied almost completely on manual contact tracing (i.e. door-to-door surveys, manual record-keeping) and monitoring to tackle earlier outbreaks, e.g., H1N1 influenza. The Integrated Disease Surveillance Program of the Ministry of Health and Family Welfare is the primary body in charge of establishing a decentralized State-laboratory based surveillance system for early detection and containment of epidemics[12]. However, realizing the importance of digital contact tracing to control the COVID-19 outbreak, the National Informatics Centre of the Ministry of Electronics and Information Technology released a smartphone-based contact tracing "Aarogya Setu" (*Bridge to Health* in English) app, which is similar to its international counterparts in that it uses Bluetooth technology to continuously monitor other devices that come within range to perform proximity tracing, and saves encrypted information of those device IDs in local memory. People who test positive for the virus are asked to manually input their test result, after which, all users who came into close contact with the patient are sent an automated alert and asked to quarantine. Aarogya Setu also uses GPS for geolocation information thus combining proximity and location tracking. The rationale given by the Indian Government for using GPS despite privacy concerns was that it could be used to quickly and more efficiently deploy resources in case of a public health emergency. Downloading the app and inputting information is mandatory for government and private sector employees and voluntary for others. Based on algorithmically defined criteria, high-risk users are identified, and their data is automatically uploaded to centralized government servers. The primary uses of the app include identification of high-risk individuals, generation of reports, creation of statistical visualizations and identification of infection clusters guide quarantine policies and accelerate decision making [13]. As of 24 April 2020, the app had 75 million downloads and while this number is high, it is still a relatively small proportion of India's approximately 500 million smartphone users[14]. A recent study (Ferretti *et al* 2020) suggested that the efficacy of digital contact tracing depends on the square of the proportion of the population using the app. This suggests that Aarogya Setu's user base has to increase significantly for it to have a meaningful effect on containment efforts. In parallel, Bharat Electronics Ltd., a public-sector undertaking under the Ministry of Defense, with expertise in networking and Internet of Things, in collaboration with the All India Institute of Medical Sciences located in Rishikesh, is developing non-invasive sensor technology for monitoring quarantined

---

[12] Vaidyanathan G 2020 People power: How India is attempting to slow the coronavirus Nature.com Online: https://www.nature.com/articles/d41586-020-01058-5 Retreived: 08 July 2020

[13] Majumdar R 2020 Coronavirus pandemic: Aarogya Setu app can help in contact tracing but privacy issues need to be addressed India Today Online: https://www.indiatoday.in/technology/features/story/coronavirus-pandemic-aarogya-setu-app-can-help-in-contact-tracing-but-privacy-issues-need-to-be-addressed-1667604-2020-04-16 Retrieved 08 July 2020

[14] The Economic Times 2020 Aarogya Setu app crosses 75 million downloads Econ. Times Online: https://economictimes.indiatimes.com/tech/internet/aarogya-setu-app-crosses-75-million-downloads/articleshow/75359890.cms Retrieved 08 July 2020

patients[15]. Measured parameters include temperature, pulse rate, peripheral oxygen saturation (SpO2) and respiration rate. These, along with the patient's location will be monitored and uploaded to cloud servers managed by a centralized command and control center set up to limit the effects of the pandemic. Sensor kits would be handed out to symptomatic patients upon assessment by clinical experts. Several HealthTech startups have also joined in efforts to create innovative solutions for curbing infection rates. Some of these have been selected by a specialized accelerator program called COVID-19 Innovations Deployment Accelerator (C-CIDA) launched by the Centre for Cellular and Molecular Platforms (C-CAMP), a bio-innovation program conceptualized by the Indian Government's Department of Biotechnology[16]. MedIoTek Health Systems (Chennai, India), one of the start-ups chosen by the accelerator program, is working on enhancing remote monitoring wearables. Their primary solution called "VinCense", monitors patients suffering from chronic heart and lung diseases and warns people nominated by the patient when an alert is triggered by the system. The company adapted its product use-case for COVID-19, stating that parameters measured by its sensors, i.e., skin temperature, respiratory rate, oxygen saturation and blood pressure, may be used for cost and time-effective triaging of at-risk individuals and in pre-screening for the virus. While there are several interesting and innovative solutions, their usefulness in supporting this pandemic remains undetermined, primarily due to a lack of government regulations for remote monitoring and lack of clinical validation. Start-ups centered around imaging technologies like Niramai Health Analytix (Bengaluru, India) and Qure.ai (Mumbai, India), have showcased the usefulness of integrating novel automated imaging methods in existing healthcare infrastructures. Niramai has adapted its patented thermal image processing technology diagnostic platform for contactless fever screening of COVID-19 patients. Similarly, Qure.ai have adapted their automated chest X-ray interpretation system (Nash *et al* 2020) originally meant for pulmonary tuberculosis screening, to monitor COVID-19 progression in the lungs. Before COVID-19, the delivery of healthcare services via digital platforms was regulated by a combination of legislative acts for information technology and medical practice[17]. Telemedicine saw a substantial rise in user engagement throughout this pandemic, with popular telemedicine platform Practo (Bengaluru, India) reporting a 5-fold increase in online consultations from March 2020 to May 2020, with 50 million users accessing their service[17]. Of these, 80% were first time users and 44% were from smaller Indian cities. In the absence of explicit government regulations, companies relied on standards defined by the United States of America's Health Insurance Portability and Accountability Act and the International Organization for Standardization for ensuring data protection and instilling trust in customers. Due to a sudden surge in the number of telemedicine users, the Ministry of Health and Family Welfare released guidelines[17] regulating the industry and establishing standards for both medical practitioners and technology platforms. This has clarified several grey areas and has stimulated more telemedicine businesses and investments in India. A bill concerning personal data protection is also currently under discussion at the Indian parliament. India was identified as the largest market for wearables in 2014[18]. A follow-up survey conducted in 2017, also identified India as one of the countries most ready for personalized monitoring devices, along with China, the US and Brazil[19]. The pandemic led to accelerated government regulations for telemonitoring and also introduced several Indians to the concept of digital health services. With greater acceptance, more start-ups are expected to propose novel solutions and methods for integrating technology with existing clinical workflows.

---

[15] Lalitha S 2020 COVID-19: BEL develops wrist, chest bands for AIIMS to monitor people under quarantine New India Express Online: https://www.newindianexpress.com/states/karnataka/2020/apr/18/covid-19-bel-develops-wrist-chest-bands-for-aiims-to-monitor-people-under-quarantine-2131943.html Retrieved 08 July 2020

[16] Fernandes S 2020 Govt, industry to support 13 innovations that may help battle Covid-19 Hindustan Times Online: https://www.hindustantimes.com/mumbai-news/govt-industry-to-support-13-innovations-that-may-help-battle-covid-19/story-WlpzQ4OfpYJ1XCIzMv95gI.html Retrieved 08 July 2020

[17] Medical Council of India 2020 Telemedicine Practice Guidelines Indian Med. Counc. (Professional Conduct. Etiquet. Ethics Regul.) Online: https://www.mohfw.gov.in/pdf/Telemedicine.pdf Retrieved 08 July 2020

*4. South Korea* Applying lessons learned from the Middle East Respiratory Syndrome outbreak in 2015, Korea has been strengthening its infectious disease surveillance and response capacity. South Korea was also an early adopter of mobile phone-assisted contact-tracing technology during COVID-19, by rolling out a government app that leverages several sources of data including cell phone location, processed CCTV footage, and credit-card records, to monitor activities[20]. When an individual tests positive, local governments can send out a general alert to all phones, that reportedly includes the individual's last name, sex, age, district of residence, and credit-card history, as well as a minute-to-minute record of their interactions with local businesses, or even with a list of rooms of a building that the person entered, when they visited a toilet, and whether they wore a mask. This approach posed severe privacy issues, and even with modifications to protect identities, the details needed to warn potential contacts of exposure risk has made it easy, particularly through the sharing of information across social media, to guess who the infectious individual is, and perhaps more importantly, what activities they had been engaged in and with whom. This has led to significant criticism of this approach, particularly in Europe and North America. In parallel, approximately one in 200 people in South Korea has been tested for COVID-19, and the country has also quickly expanded testing capacities to narrow down the infected group as soon as possible through exhaustive searching. Several hospitals have used big data information, such as GPS tracking data from phones and cars, credit card transactions, travel histories, CCTV footage, and artificial intelligence to identify high-priority cases and track the routes of infected individuals, to facilitate to implement remote diagnosis of mildly affected patients, freeing up staff to focus on diagnosing and treating severe cases. South Korea also encourages public-private partnerships to use technology to improve health results. For example, Korea Telecom has developed a global epidemic prevention platform, which has been deployed to help disinfect areas in Daegu. Hancom Group (Bundang-gu, South Korea) launched its AI Outbound Calling System to provide checkups on self-isolated citizens, asking individuals whether they have symptoms such as fever or cough, and then summarizing the statistical data, and addressing the problem of staff shortages.

*5. Japan:* Japan has a long history of promoting remote medicine, including telemonitoring and mobile clinics, with early pilots dating back to the 1970s, in an attempt to address the shortage of medical resources and regional disparities (e.g., small remote islands in rural areas). Over the past few decades, telemonitoring has proven common and useful in crucial medical areas such as

---

hypertension, diabetes, chronic obstructive pulmonary disease, heart failure, pregnancy, health promotion, and others[21]. For example, around 80% of patients diagnosed with hypertension in Japan own a home BP monitor and track the longitudinal changes in their blood pressure (BP), providing crucial information to caregivers (Anon 2014). Before COVID-19, mobile clinics in Japan were limited in use and were only utilized in rare circumstances for those in medically underserved areas. In 2018, the Ministry of Health, Labor and Welfare published the new guideline for mobile clinics[22] in which mobile clinics were no longer limited to special populations and the objectives, requirements and restrictions were clarified. The platform can either be dedicated software (e.g. Clinics[23] and Curon[24]) or common communication platforms. It is up to healthcare providers and patients to decide if they utilize these services. It is required that all services used need to be secure and privacy-protecting. The adoption of mobile clinic had been slow thanks to several restrictions in the 2018 guideline: (1) only limited numbers of diseases were covered (e.g., hypertension and diabetes); (2) "6 months rule" whereby patients must visit a doctor regularly for 6 months before they transition into telemedicine, and (3) "30 minutes rule" whereby an online healthcare provider must be located within 30 minutes from their patients in case of emergency. Although these rules were intended to promote informed decision making and prevent misdiagnosis, under treatment, misconduct, and security breaches, they hampered the rapid spread of mobile clinics. However, in response to COVID-19, these restrictions were relaxed – more diseases were approved, the leading time was shortened to 3 months (of note, during the COVID-19 pandemic, it was temporarily exemplified that a new patient could start telemedicine immediately), and the "30 minutes rule" was scrapped. Thus, these relaxations will accelerate the spread of online medicine across the country even after COVID-19. For contact tracing, Japan initially relied on voluntary phone interviews and appeared to demonstrate success in the early stage of the outbreak (Iwasaki and Grubaugh 2020). However, this approach subsequently began to lag once the numbers of new cases began to soar. On 19 June 2020, Japan launched the app COCOA (COVID-19 Contact Confirming Application), a proximity tracing app that relies on an ephemeral-ID, and thus is classified as a decentralized, privacy-preserving anonymized approach. It alerts a user if they happened to be in close vicinity (within 1 meter, >15 minutes) of someone who tested positive for COVID-19. To address privacy concerns, the program solely focuses on reporting retrospective, binary contact history within the last 14 days. Temperature monitoring is required in hospitals/clinics since Feb 25 2020 by the ministry of health labor and welfare. The program does not store, use, or report personal-identifiable information such as the user's name, phone number, and GPS information. Doctor-to-doctor communication is another emerging area in telemedicine. Intensive Care Unit (ICU) doctors whose expertise involves extracorporeal membrane oxygenation (ECMO) are scarce and are now critically needed. A team that consists of such experts provides phone consultations nationwide (dubbed as "ECMOnet"), remotely assisting physicians in deciding initiation of ECMO therapy. In summary,

---

[21] JTTA 2013 Telemedicine in Japan Japanese Telemed. Telecare Assoc. Online: http://jtta.umin.jp/pdf/telemedicine/telemedicine_in_japan_20131015-2_en.pdf Retrieved: 08 July 2020
[22] Ministry of Health Labor and Welfare 2018 オンライン診療の適切な実施に関する指針 Online: https://www.mhlw.go.jp/content/000534254.pdf Retrieved: 08 July 2020
[23] CLINICS 2020 診療業務の効率化により より良い診療を実現する Online: https://clinics-cloud.com/
[24] Curon 2020 Curon Online: https://curon.co/ Retrieved 08 July 2020 (inside Japan only)

the pandemic has, and will continue to accelerate extensive healthcare reforms that feature telemonitoring, mobile clinics, tracing apps and point-of-contact screening in Japan.

*6. Israel:* Over the last 40 years, Israel has made significant investments in many key technologies relevant to digital health. Within the scope of remote health monitoring for COVID-19, initiatives include identification of unique vocal properties of virus carriers, by the start-up Vocalis Health, working in collaboration with the Ministry of Defense and several academic groups. Their technology aims to enable mass screening and remote monitoring of recovery from COVID-19; a proof-of-concept remains to be established. The AI MyndYou startup has developed MyEleanor, which is a voice bot and virtual care manager that can act as a hotline to assess risk. TytoCare is applying its device that enables remote examination of lung and heart function and temperature, to support remote examination and monitoring of patients at home and in the hospital environment. For example, the Sheba Medical Center in Tel Aviv has been using a Tyto Care (Netanya, Israel Ltd) stethoscope and Datos Health (Ramat Gan, Israel Ltd) thermometer, blood pressure cap and CO2 device to monitor patients quarantined in a dedicated facility. Many other remote health innovations including Biobeat (Petah Tikva, Israel Ltd), EarlySense (Ramat Gan, Israel Inc) and Sweetch (Tel Aviv, Israel Ltd), have also been involved in supporting hospital and health maintenance organizations. The start-up nation has seen an important number of remote health system applications during the pandemic although the impact of these in managing the situation is yet to be assessed. Although some remote health systems (e.g. telemedicine and mobile health) were already developed in Israel prior to COVID-19, this period has provided an unprecedented opportunity for start-ups in this space to experiment with their systems and demonstrate their strength to healthcare providers such as hospitals and the Israeli health maintenance organizations. Teleconsultation services, although already existing in some form, have increased significantly during the pandemic. In particular, teleconsultation platforms were used beyond their original scope and this highlighted their shortcomings: limited functionalities, no access to patient medical records and no ability to connect to medical sensors for the purpose of remote physiological measurements. To alleviate some of these limitations many health providers made use of common teleconferencing tools such as Skype, WhatsApp or Zoom. Sometimes, patients were asked to communicate some of their medical information via email to the nurse in preparation for the consultation. The urgency of the situation thus resulted in the usage of non-secured channels of communications for delivering remote health services. It is however very much due to the urgency and need to act quickly and it is sensible to expect that more secured and dedicated platforms will emerge as a result. In the academic arena, the Artificial Intelligence in Medicine Lab at the Technion has developed a digital oximetry biomarkers toolbox that can be used both for diagnostic purposes and for monitoring pulmonary function in patients with pneumonia - a common complication of COVID-19. The toolbox is available in the open-access software PhysioZoo (Behar *et al* 2018)[25]. The Weizmann Institute of Science in Rehovot has developed a method for monitoring, identifying and predicting the zones of COVID-19 spread (Rossman *et al* 2020). To achieve that, volunteers are asked to complete a questionnaire relating to virus-induced symptoms, once a day, with the aim of identifying clusters of infection. This initiative has had tens of thousands of volunteers to date and is expected to support policymaking

---

[25] Behar J, Rosenberg A, Weiser-Bitoun I, Shemla O, Alexandra A, Konyukhov E and Yaniv Y 2020 Heart Rate Variability analysis of human and animal electrophysiological data Online: https://physiozoo.com/ Retrieved 08 July 2020

intending at slowing down the spread of the virus. Perhaps most notably, on 22 March 2020, the Israeli Ministry of Health unveiled a new location tracing app, called "The Shield" ("HaMagen")[26]. The app takes GPS location data from the user's phone and compares it with the information available within the Health Ministry servers regarding the history of the location of confirmed cases in the 14 days prior to their diagnosis. The extracted information informs users if they have crossed paths with someone who has been infected with COVID-19. The user is then given the option of reporting their exposure to the Health Ministry by filling out a form. The Health Ministry has provided public assurances that involvement is strictly voluntary and secured. The cross-referencing of the user GPS history with the ones of COVID-19 positive patients happens on the user phone and is not sent to any third party. The source code of the application has been made available to the public for transparency. Before this initiative, the movements of people diagnosed with COVID-19 were published on the Ministry's website and its Telegram channel, so identifying potential interactions with infected individuals involved many hours of manual scrolling through information. In March 2020 the Israeli government started to use geolocation data routinely collected from Israeli cellphone providers for location tracing i.e. to find people who came in close contact with identified virus carriers. Identified individuals were sent a text message directing them to isolate for 14 days. This program was adapted from a secret counterterrorism program of the Shin Bet, the Israeli domestic intelligence service, as a public health measure. The program was stopped in early June 2020 as the outbreak reduced. As of late June 2020, and while facing the second wave of COVID-19 the government decided to re-opening the program which as already sent thousands of citizen to quarentine. The attitude of Israel in that respect is "in-between" the positions of countries like China where total state surveillance is allowed and countries like France where such a program was not initiated because of the fear of intrusion on privacy. Israel chose to make use of such a program raising concerns on privacy, but it also demonstrated it can discontinue it promptly when its necessity is passed. Other initiatives from the government include the Ministry of Defense development of a monitoring system at the Laniado hospital for health professionals to remotely monitor medical data from respirator monitoring devices and the Purple Tag regulation by the Ministry of Health which is to be met by businesses willing to more fully re-open their workplace. The Purple Tag includes a set of rules that may include a tracing app where employees are asked to scan the barcode of the place they enter within the organization, filling of a daily health statement (sometimes implemented as a Google form or using a phone application) and eventual temperature checking at point-of-contact through infrared camera or portable non-contact thermometers (e.g. Figure 2). Numerous restaurants, shops and other point-of-contact places also perform a temperature check before entering. Many of the main initiatives in Israel have been led by government policy, in particular the Ministry of Health and Ministry of Defense, and financial support to selected industry (e.g. Vocalis Health Ltd. supported by the Ministry of Defense) and academic partners (e.g. Israel Science Foundation fast track COVID-19 grants).

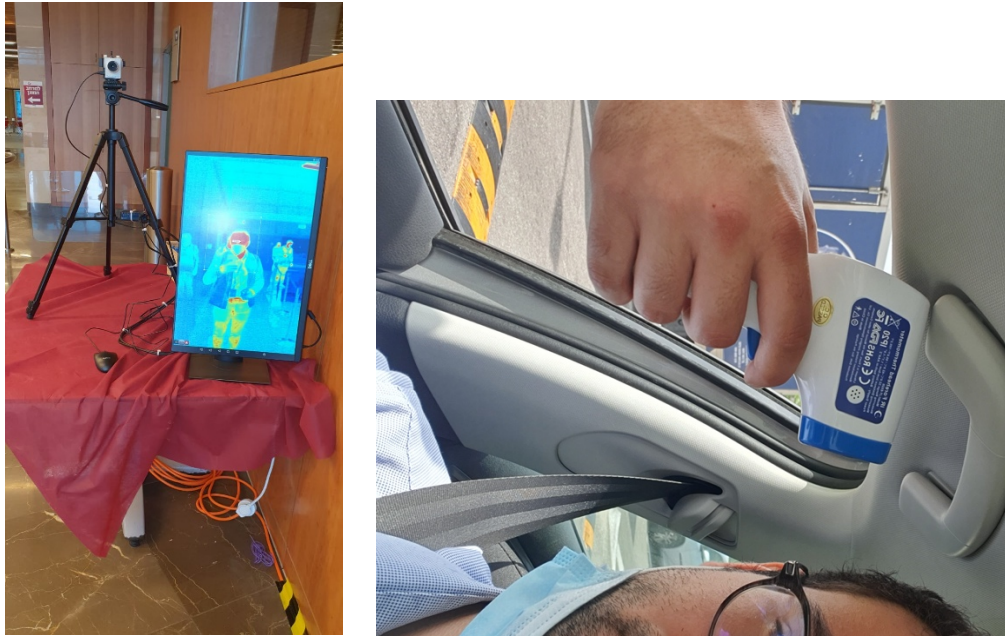

**Figure 2:** Point-of-contact temperature screening at the entrance of (left) a company in Israel and (right) at the entrance of the Technion Israel Institute of Technology in Haifa. Temperature may be checked using thermal imaging or non-contact infrared thermometers.

*7. United Kingdom:* In the UK, Nye Health (Oxford, UK), has developed a scalable bidirectional healthcare communications platform which, in theory, enables any clinician to consult any citizen, on any device from anywhere, in a privacy-preserving manner. Through its encrypted UK National Health Service (NHS) -compliant communications infrastructure, Nye Health enables both remote telephone (>90% of the remote consultation burden) and video consultations, with close to a 100% potential population penetration in any region of the UK. Before the COVID-19 outbreak, Nye Health was building "a participatory global medical community", but, by early March 2020, it redeployed its NHS-compliant infrastructure and NHS medical records integration platform in an attempt to enhance the COVID-19 response. With help from many international partners, Nye Health has since been scaling rapidly, by on-boarding clinicians and patients, and new hospitals, practices and hospices every hour and receiving regular international inquiries. Sensyne Health (Oxford, UK), developed CVm-Health, a web-based application to support individuals, families and the community to self-monitor their health during the COVID-19 pandemic. The portal allows the input of symptoms, vital signs (e.g., temperature, heart rate and blood pressure) and medication intake. Although the adoption and normalization of telemedicine has been slow in healthcare in the UK, largely due to political and organizational issues (Williams *et al* 2017), in recent weeks, COVID-19 has led the UK NHS to turn to remote consultations in order to minimize the risk of infection for staff and patients. NHS England, for example, has recruited health technology firms, such as PushDoctor (Manchester, UK), Babylon (London, UK) and Docly (London, UK), to help general practitioners set up the equipment needed for a massive expansion of telemedicine[27], either online or by phone, video or messaging services. Over the course of the pandemic,

---

[27] Campbell D 2020 GPs told to switch to digital consultations to combat Covid-19 Guard. Online: https://www.theguardian.com/world/2020/mar/06/gps-told-to-switch-to-remote-consultations-to-combat-covid-19 Retrieved 08 July 2020

universities have also been contributing to the deployment of remote vital sign monitoring solutions. A team of the Institute of Biomedical Engineering, University of Oxford, has been testing the concept of a virtual high-dependency unit, where high-risk patients are monitored in the general ward using wearable technology[28]. The system is currently being used on isolation wards in the main hospital of the Oxford University Hospitals and allows nursing staff to keep track of the condition of the patients with COVID-19 from outside the isolation rooms. It uses third-party wearables to measure vital signs (heart rate, respiratory rate, oxygen saturation and skin temperature), which are transmitted via the hospital's Wi-Fi to a dashboard. In partnership with NHS organizations in northwest London, the Imperial College of London is trialing a wearable sensor for remote monitoring[29]. Originally aimed at identifying sepsis, the wearable device is comprised of a chest patch that measures respiratory rate, heart rate and temperature every two minutes; the collected data is then analyzed by an automated algorithm, which alerts the medical team in cases of deterioration. NHSX, the technology and research arm of the NHS, developed a proximity tracing phone app with researchers from the University of Oxford and developers from tech companies like VMWare[30]. Like many other solutions, the app uses the phone's Bluetooth Low Energy network to detect and inform users about potential contact with other infected users, without identifying or disclosing personal information to other individuals. The concept is similar to that of the contact tracing app called TraceTogether, rolled out in Singapore. However, after a problematic launch on the Isle of Wight, the UK government has chosen to pivot and use a decentralized contact tracing system using an application programming interface developed for this purpose by Google (Mountainview CA, USA) and Apple Inc. (Cupertino CA, USA).

*8. France:* The practice and teaching of medicine in France has always been through physical contact with patients. For example, all medical consultation is accompanied by a physical examination and a blood pressure measurement. It is therefore hard to imagine, for French doctors as well as for patients, medical consultations without a physical examination. In July 2009, the law "Hôpital patients santé Territoires" regulated a form of remote medical practice that was made possible by technological advances. However, it remained experimental. It was only in 2016 with the establishment of the 2016 medical convention of national health insurance, that the first step enabling the reimbursement of telemedicine acts was achieved within the strict framework of the care pathway (treating doctor/patients). Concomitantly the WHO report on e-health in Europe in 2016, emphasized the interest of supporting and encouraging the deployment of telemedicine, through appropriate national health policies. Amendment 6 of this agreement, in April 2018, reinforced the deployment of telemedicine acts throughout France, by allowing their reimbursement by the national health insurance. Since 15 September 2018, teleconsultation became available throughout the French territory, as part of the coordinated care pathway [31]. However, this step forward in telemedicine was confronted with a certain "protectionism" from the medical union. For them, it was out of the question to "uberize" medical consultation. Thus, one can consider that in 2018 the legal and technological framework for telemedicine is ready but its recognition remained very fragile and subject to controversy. Its usage thus stayed very limited. The first cases of COVID-19 appeared in France at the end of January 2020 and the epidemic spread exponentially day after day, imposing on the country a total containment of the population starting on 17 March 2020. Faced with this unprecedented health crisis, calls from the "SAMU", the French emergency medical services, center 15, which is the emergency medical service in France, have increased fivefold, and medical offices were saturated. Faced with this crisis, the government (Ministry of Solidarity and Health) and the High Authority for Health (HAS) decided that teleconsultation and telecare should be offered as a first resort in order to immediately respond to the medical demand. The Minister of Solidarity and Health decided to relax the rules of practice for telemedicine on 9 March 2020, to better respond to the constraints of the epidemic. (Decree No. 2020-227). This Decree in particular included: (1) full coverage of any teleconsultation, regardless of the doctor, with a 100% reimbursement outside the treatment path; (2) the consultation must be carried out via a video exchange, and the doctor must be equipped with secure and connected equipment, in order to guarantee the security of exchanges of personal health data and the backup and protection of files; (3) possible telephone consultation is recognized during the crisis; (4) HAS files and decision tree distributed to all health stakeholders to highlight the interest and relevance of Teleconsultation. Under government influence, many start-ups and small and medium-sized enterprises members of CapDigital have mobilized to help doctors equip themselves with telemedicine and integrate regulated platforms (QARE, Doctolib, Medaviz, Livi, DocteurSécu). All medical data are protected and framed according to the General Data Protection Regulation (GDPR) of the European Union. Other specialized FrenchTech startups provide data protection solutions so that no French doctor needs to use non-secure tools like WhatsApp, FaceTime, Skype, or Zoom. The confinement significantly increased the usage of telemedicine, with 1 million teleconsultations recorded by Caisse Nationale de l'Assurance Maladie (CNAM) or the week of 30 March 2020 alone. During a press release from the national health insurance on the 31 March 2020, teleconsultations were now constituting more than 11% of all consultations, compared to less than 1% before the crisis. On 7 April 2020, the CNAM listed 30,000 doctors had adopted teleconsultation regularly. More than 44% of general practitioners having experimented using teleconsultation. The pandemic has, therefore, played the role of an accelerator, making it possible to overcome within a few days many obstacles that were limiting the development of remote health monitoring in France. Unlike other countries like Israel and China, France has not implemented a patient tracking policy using mobile phone geolocation. The government has said it is in favor of digital tracking of the French, on a voluntary basis, to fight COVID-19 with the use of the app StopCovid. StopCovid is an application that is part of the government's deconfinement plan. The stated objectives of StopCovid are for citizens to protect themselves, protect others, and support the efforts of caregivers and the health system to quickly stop the chains of contamination of the Covid-19 epidemic. On the 27 May 2020, the National Assembly and the Senate voted in favor of the deployment of StopCovid. The StopCovid application has been available on the Apple Store and Google Play since 2 June 2020. The application has been downloaded over 1.9 million times since its launch, for a total of 1.8 million activations in June 2020. Teleconsultation, alternating with face-to-face consultation, makes it possible to simplify the monitoring of chronic pathologies by limiting the movement of patients.

---

[28] Lyons E 2020 John Radcliffe coronavirus patients monitored with wearable tech Oxford Mail Online: https://www.oxfordmail.co.uk/news/18361997.john-radcliffe-coronavirus-patients-monitored-wearable-tech/ Retrieved 08 July 2020

[29] Alford J 2020 Wearable sensor trialled for remote COVID-19 monitoring Imp. Coll. London News Online: https://www.imperial.ac.uk/news/196973/wearable-sensor-trialled-remote-covid-19-monitoring/ Retrieved 08 July 2020

[30] Wright M 2020 Contact tracing app could halt spread of COVID-19 if downloaded by 60 per cent of the population Telegr. Online: https://www.telegraph.co.uk/news/2020/04/26/contact-tracing-app-could-halt-spread-covid-19-downloaded-60/ Retrieved 08 July 2020

[31] Maladie C nationale de l'Assurance 2018 GENERALISATION DE LA TELECONSULTATION LE 15 SEPTEMBRE 2018 Online: https://www.ameli.fr/fileadmin/user_upload/documents/Dossier-de-presse_Teleconsultation_12092018.pdf Retrieved 08 July 2020

This also makes it possible to provide faster care and support in the event of an emergency and acute seasonal pathology (winter pathology). Finally, in the government's "ma santé 2022" plan[32], telemedicine is cited as an effective means against these geographic inequalities of access to care; 8 million French people live in a medical desert. Overall, the pandemic had a dramatic impact on the way medical services are delivered in France. Moving from a traditional and very protectionist stand, medical professionals and patients were pushed to experiment with telemedicine platforms. In time the benefit that these remote services have on the medical system in France will be understood, and this will have an impact on the future of the French healthcare system. Indeed, even with the pandemic, the French centuries-long traditional practice of medicine will take some time to change.

*9. Italy:* A European Union (EU) study[33] analyzed the market for telemedicine applications and solutions. The study showed that Italy makes up a large proportion of the EU telehealth market, with a telehealth market revenue per capita of 3.38 Euros per annum (in comparison, the highest spent per capita is Denmark with 6.22 Euros). In Italy, electronic prescriptions, considered as an indicator of the uptake of telemedicine tools, are only issued by 12% of the general practitioners. While some telemedicine services today are eligible for reimbursement, patients still bear most of their costs. Telemedicine has been integrated as an option in standard home hospitalization and integrated home care. However, these efforts have been rather fruitless since the reimbursement schemes and financing structures have not kept pace with these changes. The implementation of telemedicine in the field of cardiovascular disease was recently assessed by a census, supported by the Italian Society of Cardiology (Brunetti *et al* 2020), which showed that the principal area of interest in telemedicine for cardiovascular disease is prehospital triage of acute myocardial infarction. Another cluster of telemedicine activity is dedicated to outpatients with chronic cardiovascular disease, to enable easy access to cardiology examinations such as electrocardiograms (ECGs), ambulatory ECG or blood pressure monitoring. Another example of telemedicine in Italy is the screening of diabetic retinopathy (Vujosevic and Midena 2016), for the management of elderly complex patients with one or more comorbidities (Scalvini *et al* 2018). However, Italy does not include telemedicine in the essential levels of care granted to all citizens within the Italian National Health Service. The lack of reimbursement has been flagged as the main barrier toward the implementation of remote health monitoring (Palmisano *et al* 2020). At the beginning of the COVID-19 emergency, no formal input on telemedicine was given by health authorities, despite the high pressure on health services during the first phase of the epidemic. Later, an open call for telemedicine and monitoring system technologies proposals was jointly issued on 24 March 2020 by the Ministry for Technological Innovation and Digitalization, the Ministry of Health, the National Institute of Health and the WHO[34]. However, currently, the only outcome of this call was the announced implementation of a smartphone app to be used for contact tracing. As yet, no large-scale telemedicine services for monitoring acute and chronic patients and allowing continuity of care have been considered. Omboni (Omboni 2020) has offered several explanations as to why Italy has missed a unique opportunity to set up an infrastructure for providing care via telemedicine and to usher in a care system. First, the available solutions do not allow the integration of the available systems with the electronic health record of the national health system. Second, in most cases, there is poor interconnection between telemedicine services operating at higher levels (secondary or tertiary care facilities) and those deployed in primary care clinics or community pharmacies. Third, many telemedicine services funded by the local or central governmental institutions lack evidence for clinical and cost-effectiveness. Fourth, the implementation of telemedicine solutions is often hampered by heavy privacy regulations and the lack of practical recommendations. Finally, telemedicine services are not yet included in the basic public health care packages granted to all Italian citizens. A report with interim indications for telemedicine assistance services was drawn up by the Istituto Superiore di Sanità (Gabbrielli *et al* 2020). The report emphasizes that the recommendations only apply during the COVID-19 emergency and are not to be continued afterwards. The listed indications are designed to offer health services and psychological support to people in quarantine at home, to proactively monitor their health conditions. The aim of a home telemedicine service is to bring medical services to people in isolation or isolated following the rules of social distancing, in order to proactively monitor their health conditions, in relation both to the prevention and treatment of COVID-19 and to ensure the continuity of care that may be necessary for other pathologies and/or conditions that require it. Video examination calls have been established, being active tele-control of the health: (1) for people who were in contact with a confirmed COVID-19 case to detect the possible appearance of signs and symptoms of COVID-19 infection; (2) for the necessary treatments against COVID-19 and to arrange for any hospitalization, when appropriate; and (3) to provide the best possible continuity of at-home care and assistance, in relation to the patient's basic condition and any COVID-19 infection. For all examinations, a list of symptoms to be checked is provided, some requiring specific devices, such as a digital pulse oximeter or ECG recorder, but there is no mention of alternatives in cases in which the person does not have such devices at hand. Practically speaking, the implementation of telemedicine in Italy during the current COVID-19 pandemic has been mainly limited to the possibility of fully digital prescriptions and some routine examinations. In summary, the adoption of telehealth in Italy has been very low, likely due to the lack of reimbursement schemes. With the escalation of the current pandemic, specific remote medical services have become available. There have been very few technological developments both from start-ups and the government, in the field of remote sensing technology. On 15 June 2020, the app Immuni[35] created by the start-up Bendin Spoons S.p.A. but managed by the Government for contact tracing was made available and active throughout Italy. Privacy issues have been taken into account (as little personal data as possible is recorded) and it is based on Bluetooth Low Energy and not on GPS location tracing. The app was made available on 15 June 2020, and after two weeks, only 4 million Italian citizens had downloaded the app, not enough yet for reliable contact tracing. After lockdown temperature checks will be performed before entering public places. The post-COVID-19 era will likely see the healthcare system slip back to where they were before, except, for some remote services that may hopefully remain.

---

[32] Ministres des Solidarités et de la Santé. (2018). *Ma santé 2022 : un engagement collectif*. https://solidarites-sante.gouv.fr/IMG/pdf/ma_sante_2022_pages_vdef_.pdf Retrieved 08 July 2020

[33] PWC for European COmmission 2018 Market study on telemedicine *Third EU Heal. Program.* Online: https://ec.europa.eu/health/sites/health/files/ehealth/docs/2018_provision_marketstudy_telemedicine_en.pdf Retrieved 08 July 2020

[34] AGID 2020 La tecnologia e l'innovazione per la lotta al Coronavirus. Le fast call Innova per l'Italia Online: https://innovaperlitalia.agid.gov.it/call2action/index.html Retrieved 08 July 2020

---

[35] Immuni immuni Home Page Online: https://www.immuni.italia.it/ Retrieved 08 July 2020

*10. Germany:* While Germany has pursued remote health care projects for decades (Stroetmann and Erkert 1999), current practices barely integrate them (Brauns and Loos 2015). The lack of reimbursement programs, as well as legal restrictions, e.g., the prohibition of remote treatment, have posed near-insurmountable barriers when trying to establish innovative solutions for remote healthcare in Germany. Acknowledging this situation, the German Parliament passed the digital healthcare act (Digitale Versorgung Gesetz, DVG) in 2019, i.e., just before the COVID-19 outbreak. The DVG established a revised legal framework and has taken actions, e.g., the mandatory launch of electronic health records, i.e. an electronic collection of personal data, past and present illness and treatments, the authorization of video consultations and changes in reimbursement[36], to promote remote healthcare. Importantly concerning future remote health monitoring, the DVG enables so-called digital health applications (Digitale Gesundheitsanwendungen, DiGA)[37], i.e., products that capture data, monitor or treat patients, which can be used by patients or patients together with healthcare professionals. The experiences from COVID-19 and its severe country-wide consequences, together with the novel legal situation, may act as game-changer in the remote health arena in Germany. COVID-19 has immediately triggered solutions for remote healthcare, e.g. informative apps on COVID-19, apps for personal risk assessment and an app, which gathers data (e.g. movement pattern, heart rate, sleep behavior) by a smartphone and a connected wearable to gain a better understanding on the spread of COVID-19 and provide information on it to the public[38]. Many of the pursued approaches bear large potential, but only a few applications managed to reach the broad public quickly. Maybe the most important success story for telehealth is the use of teleconsultations, which experienced a sudden increase within the pandemic. According to an estimate of the National Association of Statutory Health Insurance Physicians, the number of doctors' practices that offer video consultation raised from 1700 in February 2020 to 25000 in April 2020[39]. A recent survey among German doctors revealed that 84 % of doctors who offered video consultations before the pandemic practically did not practice them until the pandemic. During the pandemic, however, 95% of them practiced video consultations[40]. Here, the already modified legal framework and the urgent need for social distancing had an immediate and strong effect. The surveyed doctors further assumed video consultations to retain relevance even after the pandemic, which reflects changes in doctors' attitudes towards the use of remote technology because of COVID-19 and a sustainable effect. Short to medium term, it is thus likely that even other telehealth services will gain importance. Specific funding which was and will be launched by the government might support this development. However, privacy and data protection area are of major concern among Germans. A tracing app, i.e. an app commonly used to track infections and contacts to infected persons, can serve as an illustrative example; although practical evidence and simulated data (Kucharski *et al* 2020) confirm the benefit of a tracing app

---

[36] Federal Ministry of Health 2019 Driving the digital transformation of Germany's healthcare system for the good of patients. The Act to Improve Healthcare Provision through Digitalisation and Innovation Fed. Minist. Heal. Online: https://www.bundesgesundheitsministerium.de/digital-healthcare-act.html Retrieved 08 July 2020

[37] Federal Institute for Drugs and Medical Devices 2020 Digital Health Applications (DiGA) Fed. Inst. Drugs Med. Devices Online: https://www.bfarm.de/EN/MedicalDevices/DiGA/_node.html Retrieved 08 July 2020

[38] Corona-Datenspende 2020 Corona-Datenspende Robert Koch Inst. Online: https://corona-datenspende.de/

[39] Pritzkow K 2020 Telemedicine: the clinic of the future? Mediterr. Broadcast. Online: https://www.mdr.de/nachrichten/ratgeber/gesundheit/anstieg-nutzung-telemedizin-chancen-und-huerden-100.html Retrieved 08 July 2020

[40] Obermann K 2020 Ärztliche Arbeit und Nutzung von Videosprechstunden während der Covid-19-Pandemie Online: https://www.stiftung-gesundheit.de/pdf/studien/aerzte-im-zukunftsmarkt-gesundheit_2020.pdf Retrieved 08 July 2020

as part of a multifactorial strategy to slow down the spread of COVID-19 it has been slow in taking off in Germany. Indeed, the official app initiated by the German government (intended for voluntary use) was released mid-June 2020, i.e. more than four months after intense discussions on it started, most importantly owing to the discussion regarding the pursuit of a solution with central versus decentralized data processing, finally opting for the latter.

*11. Switzerland:* Compared to Germany, Switzerland had telemedicine services that were already well established and promoted by health insurers before the COVID-19 outbreak. With the spread of the disease, these services have received strong support from the population and consequently, their use increased by 30% by April 2020[41]. However, these services are largely based on videoconferencing and do not support remote vital sign measurements. In the wake of COVID-19, several medtech startups have launched initiatives to bridge this gap. Leitwert GmbH (Zurich, CH) has partnered with the University Hospital Basel (Basel, CH) to remotely track vital signs with wearables, which are continuously stored using their internet of medical things device and database management solution. Third-party wearables can be easily integrated to transmit health parameters independent from location to a cloud-hosted dashboard (Käch *et al* 2020). Ava AG (Zurich, CH) has made available their Ava bracelet, which was originally designed for pregnancy planning, to monitor patients with COVID-19 symptoms (Goodale *et al* 2019). The bracelet is being distributed as a research and early surveillance tool in a large-scale study involving over 2000 subjects in Lichtenstein. CSEM SA, a Neuchâtel based, private, non-profit Swiss research and technology organization that also specializes in wearable electronics, is partnering with various hospitals in and outside the country to monitor COVID-19-positive heart rate, temperature, respiration and oxygen saturation. One of the major initiatives in the remote monitoring domain evolved at the two engineering universities EPFL (Lausanne, CH) and ETH Zurich (Zurich, CH). They developed, along with other European partners, the open-source Decentralized Privacy-Preserving Proximity Tracing (DP-3T) Protocol (Troncoso *et al* 2020). The DP-3T is a Bluetooth-based mobile phone sensing approach to automatically track and inform users about potential contact with other users who self-reported a SARS-CoV-2 infection. The decentralized approach stores anonymized proximity patterns locally on the phone which maximizes privacy for the individual, without disclosing identities or other personal information and connections to other individuals or governments. The SwissCovid App is a proximity tracing solution based on DP-3T and the corresponding iOS and Android interfaces. It entered public beta testing on 25 May 2020 and is worldwide the first based on this technology. After approval and legislation change from the Swiss parliament on 8 June 2020, it was launched by the Swiss Federal Office of Public Health across Switzerland on 25 June 2020. At the time of writing, the Swiss Association of Research Ethics Committees (Swissethics, Bern, CH) have reported on four approved research studies involving telemonitoring or point-of-contact screening, and an additional four were under evaluation[42].

*12. Portugal:* Portugal can also be considered an early adopter of telemedicine, in pursuit of ways to overcome its geographic barriers and shortage of professionals (Castela *et al* 2005, LV *et al*

---

[41] Möller V 2020 Videosprechstunden: Was in der Schweiz längst normal ist, boomt in Deutschland durch die Corona-Krise *Reue Burcher Beitung* Online: https://www.nzz.ch/wirtschaft/corona-krise-telemedizin-in-deutschland-und-der-schweiz-boomt-ld.1550704?reduced=true Retrieved: 08 July 2020

[42] Schweizer Ethikkommissionen 2020 List of approved clinical trials and research projects on COVID-19 in Switzerland. Swiss Assoc. Res. Ethics Committees Online: https://swissethics.ch/covid-19/approved-projects Retrieved 25 June 2020

n.d.). Portugal has been implementing telehealth since the 1990s and has led many initiatives that have been adopted throughout the country in the last two decades. Examples of these initiatives include real-time remote consultations between healthcare providers, for instance, between primary care physicians and hospitals (Maia *et al* 2019), remote screening, particularly in the area of dermatology in rural areas (Oliveira *et al* 2014), and remote monitoring programs for chronic diseases, such as chronic obstructive pulmonary diseases and chronic heart failure[43]. These already standard information and communication technology systems typically support video-calls, including the sharing of clinical images or lab tests during the sessions. The interest of the country in these technologies is well demonstrated by the newly introduced National Strategic Telehealth Plan launched by the Portuguese Ministry of Health in November 2019, just before the new COVID-19 outbreak, to advance telehealth. It includes input from 50 institutional stakeholders and experts across the health spectrum, and describes strategies that will drive improvements in infrastructure, interoperability, and serviceability, and will support the required legislation. The Portuguese National Health Service already provides a whole suite of digital tools to help people manage their health remotely[44]. On the online Citizen Area (RSE Área do Cidadão)[45] for example, users can access their electronic health record (EHR), book an appointment with their primary care physician and check their vaccination card. A particularly powerful service is compulsory within the Portuguese National Health Service (SNS) and used by private healthcare patients. Additionally, the MySNS mobile app[46] is a wallet where users can gather personal information about their healthcare such as vaccine cards, access data to the service (SNS), allergy registration and e-prescriptions. The app can also create reminders on smartphone calendars, reminding patients, for example, when to take their medication. Despite the difficulty in quantifying the impact of COVID-19 on the usage of these tools for remote monitoring, early indicators suggest an acceleration on their adoption; e.g., in March 2020, when lockdown measures were first implemented due to COVID-19, the number of telemedicine consultations increased by 44% as compared to March 2019[47]. Portuguese startups started to offer telemedicine services free of charge to help the SNS manage cases of COVID-19. As a summary, there have been some advances in implementing and deploying telemedicine systems in Portugal in the past decades, which have been transforming the country's health care system. During the evolving pandemic, these systems have been further used and novel telehealth technologies for the remote monitoring of different conditions may be developed and become available to the population.

*13. Russia:* In Russia, the market of telehealth and remote health monitoring is still in early stages. The regulatory system did not authorize telehealth services until 29 July 2017, when Federal Law №242 "On introducing changes into selected regulations of Russian Federation on the topic of using information technologies in healthcare" was passed, allowing for the provision of

---

[43] SPMS 20016 Comissão de Acompanhamento da Informatização Clínica – Relatório de Atividades, Lisboa Serviços Partilhados do Ministério da Saúde Online: http://www.cnts.min-saude.pt/2017/03/28/211/ Retrieved 08 July 2020

[44] SNS 2017 Citizen's area – SNS Portal: More than 5000 daily appointments scheduled on the online platform Serviço Nac. Saúde Online: https://www.sns.gov.pt/noticias/2017/10/06/area-do-cidadao-portal-sns/ Retrieved 08 July 2020

[45] SNS 2020a Área do Cidadão do Portal SNS Serviço Nac. Saúde Online: https://servicos.min-saude.pt/utente/

[46] SNS 2020c MySNS Carteira Serviço Nac. Saúde Online: https://www.sns.gov.pt/apps/mysns-carteira-eletronica-da-saude/ Retrieved 08 July 2020

[47] SNS 2020b Consultas em Telemedicina Serviço Nac. Saúde Online: https://www.sns.gov.pt/monitorizacao-do-sns/consultas-em-telemedicina/ Retrieved 08 July 2020

teleconsultation (excluding diagnosis) and electronic prescription services, starting 2018[48]. This change in legislation prompted the arrival of a wave of telehealth startups, mostly focusing on low-hanging fruits of remote consultations and using a marketplace business model. Following the initial enthusiasm and rapid technology development, a market analysis by EY from early 2020 found that 81% of private-sector healthcare organizations that invested in telemedicine services following the legislation change, considered these projects unsuccessful due to low adoption and resistance of both patients and clinical personnel[49]. Yet, the SARS-CoV-2 pandemic has reversed this trend again, with remote consultations and electronic prescriptions becoming state-funded and provided services, following the recommendations of the Ministry of Health. Electronic prescriptions have also become the recommended approach for chronic disorders. As state regulation plays the most significant role in shaping the telehealth market in Russia, where only 10-15% of major city inhabitants use private healthcare, telehealth may become a de-facto standard of healthcare service due to the change of public perception and rapid introduction of new technologies during the current pandemic. Several commercial telehealth services like Doc+[50], DocDoc[51], OnlineDoctor[52], Telemed24[53] and Yandex.Health[54] were already in operation when the SARS-CoV-2 pandemics started. All of the services provide a moderate level of personal data protection. All of them require account registration with a track of patients' medical records. At the moment all the services report an increase of telehealth appointments and expect a doubling of the appointments by the end of 2020. A new general telehealth service Doctor Near [55] (Доктор рядом in Russian) was launched in March 2020 at the beginning of the COVID-19 outbreak in Russia. This is a startup backed by a state-owned bank. The service reports about 60,000 telehealth appointments in 2 months 70% of those for acute respiratory infection[56]. The service requires registration and track medical records of the patients to provide a moderate level of personal data protection. All these services apply telemedicine tools to diagnose and manage COVID-19 patients as well as the management of other acute and chronic conditions at distance. Moscow regional department of health has organized a telehealth consultation center for the ambulatory patients diagnosed with SARS-CoV-2. The patient has the opportunity to make an individual appointment for a consultation, but the doctor can also plan an online appointment and invite a patient. The telemedicine center works in 24/7 mode to consult patients around the clock. The capacity of the Center is 200 specialists, which is enough to conduct up to 4000 consultations per day. One appointment lasts on average 13-15 minutes. By 15 June 2020, the service has provided over 340,000 remote consultations for 114,000 patients with COVID-19. The service provides a moderate data protection level. Doctors have two screens: one for the telehealth service and another shows the electronic health record from the United Medical

---

[48] Russian Federation 2017 О внесении изменений в отдельные законодательные акты Российской Федерации по вопросам применения информационных технологий в сфере охраны здоровья, федеральный закон № 242-ФЗ Online: http://publication.pravo.gov.ru/Document/View/0001201707300032 Retrieved 08 July 2020

[49] EY 2020 Исследование рынка коммерческой медицины в России. 2018-2019 годы. Market Report Online: https://assets.ey.com/content/dam/ey-sites/ey-com/ru_ru/news/2020/03/ey_healthcare_research_2018-2019_24032020.pdf

[50] Doc+ Telemedicine in Russia Online: https://docplus.ru/services/telemedetsina/ Retrieved 08 July 2020

[51] DocDoc Найдите проверенного врача и запишитесь на приём Online: https://docdoc.ru/ Retrieved 08 July 2020

[52] LLC Mobile Medical Technologies Online Doctors Home Page Online: https://onlinedoctor.ru/doctors/ Retrieved 08 July 2020

[53] О ТелеМед О ТелеМед Home Page Online: https://www.telmed24.ru/ Retrieved 08 July 2020

[54] Клинике Яндекс.Здоровье Онлайн-консультации с врачами Online: https://health.yandex.ru/ Retrieved 08 July 2020

[55] Anon Официальный сайт государственной программы «Доктор рядом» Online: http://www.dr-clinics.ru/# Retrieved 08 July 2020

[56] Anon Болезнь попалась в сети Online: https://novayagazeta.ru/articles/2020/05/29/85601-bolezn-popalas-v-seti Retrieved 08 July 2020

Information and Analytical System of Moscow [57]. Symptom checkers are also gaining momentum during SARS-CoV-2 pandemics in Russia. The first symptom checker to provide SARS-CoV-2 specific dialogue was helzy.ru from Medlinx LLC. This is an industry initiative to provide self-diagnostics and routing to a relevant doctor. The service is completely anonymous and provides a strong data protection level. It does not collect or store medical data. The symptom checker was available since late 2019 and according to the developers gained around an additional 30% of users with the emergence of COVID-19 in Russia. "Social monitoring" is an initiative of a Moscow government to track that confirmed COVID-19 patients or their home mates follow the prescribed quarantine obligations[58]. The mobile application "Social Monitoring" is obliged to be installed by all people who have confirmed COVID-19 infection, but they are being treated at home in mandatory isolation. The service is mandatory for Moscow residents with a confirmed diagnosis of COVID-19, who are under treatment at home; patients with symptoms of acute respiratory disease (ARD) and those who live with patients with confirmed COVID-19 or ARD. The Social Monitor app records the location of the phone and sends a selfie photo request several times a day to check that the user is nearby. If the selfie photo does not reach the server of the Moscow Main Control Department within an hour, the person will be issued an administrative fine. By the end of May 2020, there have been around 67000 active users. The service has a little protection level that is still in compliance with Russian data protection regulations (FZ 152)[59]. The personal data that is provided to the service is defined in the consent to receive medical care at home or by a decree of the chief medical officer. This includes a surname, first name, patronymic name, the address at which the patient decided to undergo treatment and a mobile phone number. In addition, by signing these documents, you give your consent for taking photographs and presenting an identity document, as well as for processing the personal data. The data that the user sends to the application are stored on the servers of the Information Technology Department in a secure form. The service was introduced entirely for the purpose of monitoring of COVID-19 quarantine in Moscow. Other regions of Russia did not report the implementation of COVID-19 specific telemonitoring solutions or specializing existing telemedicine services for COVID-19.

*14. Canada:* In the face of the COVID-19 pandemic, several Canadian initiatives are underway to provide innovative, urgently needed technological support for remote health monitoring. Contact tracing, perinatal surveillance, remote ICU telemonitoring and virtual critical care are just a few important examples discussed below. AI researchers led by Yoshua Bengio, from the MILA Institute, a partnership between the Université de Montréal and McGill University, have partnered with Quebec entrepreneurs to develop a peer-to-peer AI-based platform for COVID-19 tracing[60]. Rather than identifying individuals (which could lead to stigmatization), the goal is to provide citizens with information to reduce their risk of viral contamination. The key privacy element is that all information is kept strictly anonymous, and a subset of the data is used by a centralized "data trust" to periodically retrain a risk predictor. This proximity-based tracing intends to "apply social distancing at the right places, around infected people". As with many of

---

[57] ЕМИАС Запись к врачу в городские поликлиники Москвы Online: https://emias.info/about Retrieved 08 July 2020
[58] mos.ru Социальный мониторинг Online: https://www.mos.ru/city/projects/monitoring/ Retrieved 08 July 2020
[59] ФЕДЕРАЛЬНАЯ СЛУЖБА ПО НАДЗОРУ В СФЕРЕ СВЯЗИ И Т И М К Federal Law of 27 July 2006 N 152-FZ ON PERSONAL DATA Online: https://pd.rkn.gov.ru/authority/p146/p164/ Retrieved 08 July 2020
[60] Yoshua Bengio 2020 Peer-to-peer AI-tracing of COVID-19 Online: https://yoshuabengio.org/2020/03/23/peer-to-peer-ai-tracing-of-covid-19/ Retrieved 08 July 2020

the contact tracing apps, the idea is to have a smartphone indicate one's current probability of infection conditioned by where they have been and who has been in their proximity, from information gathered by Bluetooth peer-to-peer communication. That risk is then shared with those they subsequently encounter. Following a test for infection, medical personnel use encrypted communication to provide the results to the patient's smartphone application and to upload anonymized and delocalized patient information (without using geographical trajectories, but, rather, sequences of encounters and associated risks) from their smartphone to a non-governmental data trust that collects data to train the risk predictor application, whose updates are periodically disseminated to participants who provide data. Rapid deployment of this application requires the collaboration of the provincial authorities (the Quebec Ministry of Health in this case) that administer the Canadian public health system; such negotiations are ongoing. Other contact-tracing initiatives are also in place in several other provincial jurisdictions. Toronto's University Health Network has adapted a telemonitoring system for COVID-19 patients, to enable remote observation by nursing staff. Originally designed to ensure that lung surgery patients maintain their oxygen masks in position, as the pandemic arose, rapid and cost-effective solutions to reduce COVID-19 exposure were required for attending nurses, whose PPE was quickly in short supply. The system, now in operation for hospitalized COVID-19 cases, enables remote monitoring by the nurse, using one camera to observe the patient and another to monitor medical devices measuring blood pressure and oxygen saturation. Audio communication allows the nursing staff to interact with the patient and remind those who are capable to reposition the oxygen probe when necessary. If oxygen levels fall to critical levels, a sentinel event in deteriorating COVID-19 patients, physical intervention is initiated. Remote perinatal surveillance allows obstetrical staff to telemonitor mothers in the delivery ward from a central nursing station or from other facilities. Networked cardiotocography (CTG) and automated recognition of CTG signal patterns (fetal heart rate decelerations/accelerations, uterine contractions) (Warrick *et al* 2005), such as those provided by PeriGen Inc. (Montreal, Quebec and Cary, North Carolina) and by GE HealthCare (Chicago, IL) allow nurses to care for multiple patients at once and especially during the pandemic, can reduce exposure of clinical staff to infection over potentially many hours of labor. It can also provide support for lesser-experienced nurses who have been reassigned during the crisis from other departments. These CTG resources make better use of obstetrical expertise that is in short supply, increasing the pool of obstetricians who can observe the recordings and discuss the degree of concern about the progress of labor and the fetal state. Care in close proximity to the mother is still required for periodic pelvic exams and at the time of delivery. As COVID-19 infection spread in Canada, construction of makeshift hospitals began[61] and teleconsultation replaced a significant proportion of the in-person patient-clinician encounters. But for some time, with Canada's large geographical expanses, virtual critical care has been an important tool to provide family medicine in outlying clinics with videoconferencing access to specialists at tertiary care centers, including critical care (ICU) physicians and nurses, pharmacists and respiratory therapists. Such a model enables patients who are critically ill to be cared for locally when possible. The COVID-19 context has amplified the need for these services, which is expected to continue for some time as the virus continues to spread to more remote regions. One such health network expanding its remote services in the context of COVID-19 is the Health

---

[61] The Globe and Mail, Ontario, B.C., Quebec begin building makeshift hospitals in preparation for rise in COVID-19 patients, Online: https://www.theglobeandmail.com/canada/article-ontario-bc-quebec-begin-building-makeshift-hospitals-in/ Retrieved 08 July 2020

Sciences North Intensive Care Unit in Sudbury Ontario, which is a hub for remote consultation services, providing support for numerous Northern Ontario communities, such as the far-north nursing station at Wayneebayko Health Authority in the Indigenous Cree communities of the James Bay Lowlands. The McGill Integrated University Health Network (RUIS McGill) is a similar telemedicine system that have been in place since 2006 to service Quebec communities stretching from Montreal to Nunavik.

*15. United States:* In the United States, in response to the demand for flexibility and broadened access to telemedicine services, changes by the Health and Human Services (HHS) Office for Civil Rights (OCR) and the Centers for Medicare & Medicaid Services (CMS) have led to a relaxation of some federal privacy regulations and expansion of payment policies[62]. This has led to a surge in the supply of tele-consults around the US. However, very few, if any, new devices are being considered, and traditional FDA-approved devices are still required. Several online symptom checkers have been published to reduce unnecessary emergency department visits[63,64] although the empirical evidence for these remains unclear. While they vary in logical flow, they tend to agree on the key risk factors. At the same time, it is important to note that the FDA is fast-tracking approvals for COVID-19-related diagnostics or screening devices and relaxing restrictions on the usage of certain devices as well[65]. For example, Alivecor is currently fast-tracking a QT screening algorithm for the Kardia 6L remote monitoring electrocardiogram (ECG) device. Multiple smartwatch/fitness tracker studies have also been initiated across the US. Stanford University and Scripps Research Translational Institute have partnered to harvest data from Fitbits, Apple Watches, and other smartwatches/fitness trackers to enable "real-time surveillance of contagious respiratory illnesses." These avenues are predicated on the assumption that fever is associated with a persistent rise in heart rate. However, because heart rate depends on so many factors, it's not entirely clear that this would work with any reasonable level of specificity. Moreover, sick people tend not to wear or charge personal tracking devices with any regularity, or (it is probably safe to assume) when in lockdown. UC San Francisco has provided healthcare workers with *Oura* rings, worn on the finger, which track heart rate, peripheral temperature, and nighttime respiratory rate. However, even though temperature is measured, correlation with core temperature is likely to be poor. Indeed, core temperature is particularly hard to identify, with far-infrared (FIR) thermal imaging being a contender for remote/non-contact monitoring. A team at Emory University and the Georgia Institute of Technology (Hegde et al. 2020) recently developed the "AutoTriage" COVID-19 triage system for rapid screening at the emergency department or health posts. The system involves a visible and FIR camera mounted on a Raspberry Pi mini-computer. Deep learning algorithms running on a dedicated USB tensor processing unit, identify the forehead and lips (potentially on up to ten people simultaneously) and co-registers the image from the visible to FIR. Temperature is read from the forehead in the visible domain and cyanosis is determined from the lips in the visible domain. Heart rate and respiration activity extraction are currently being added. The entire code stack code and bill of materials was published under an open-source license and is undergoing live trials to determine if and how ambient light and temperature affect the system. However, significant work is needed before this system will be ready for remote monitoring. As noted, the most obvious remote sensing system already in our hands is the modern smartphone. Studies have shown that activity associated with a smartphone can help detect or predict changes in health, including sleep apnea (Behar *et al* 2015), Parkinson's disease (Neto *et al* 2016) heart failure (Cakmak *et al* 2018) and several other conditions (Trifan *et al* 2019). Many research groups, spanning from the east (*safepaths.mit.edu*) to the west (*Covid Watch*) coast of the USA, have implemented contact tracing apps with a strong focus on privacy. Safe Paths describes itself as "an MIT-led, free, open-source technology that enables jurisdictions and individuals to maximize privacy, while also maximizing the effectiveness of contact tracing in the case of a positive diagnosis". The Safe Paths platform, currently in beta, comprises both a smartphone application, PrivateKit, and a web application, Safe Places. The PrivateKit app will enable users to match the personal diary of location data on their smartphones with anonymized, redacted, and blurred location history of infected patients. The digital contact tracing uses overlapping GPS and Bluetooth trails that allow an individual to check if they have crossed paths with someone who was later diagnosed positive for the virus. Through Safe Places, public health officials can redact location trails of diagnosed carriers and then broadcast location information, with privacy protection, for both diagnosed patients and for local businesses." Stanford's COVID Watch, also uses Bluetooth to determine if a user's phone approaches within approximately six feet of another smartphone that also has the app installed and running, and "maintains" that proximity for 15 minutes or more, during which, the two phones share a temporary contact number that is stored on each device. In this manner, no data ever leave the phone (unless the user actively sends out information), and the data that are stored locally are anonymous. If an app user is confirmed positive, they can send their anonymous personal identifier data to a cloud storage repository. The app will then alert other app users who spent 15 minutes or more near the infected person. In this way, people can choose to follow the local guidance, such as calling their public health department, without the fear of being tracked or contacted directly. COVID Watch relies on self-reporting, and the developers assume that no responsible person, knowing they were COVID-positive, would purposefully risk spreading the disease through prolonged contact with others. These approaches contrast heavily with the Israeli, South Korean, Chinese and Australian approaches of a government-managed technology, and reflect the deep distrust residents of the USA have of their government. There are two obvious flaws here. First, the apps require a critical mass of users on the same platform, to allow the app to run constantly. A proliferation of multiple apps that do not communicate with each other erodes the utility of the app. Second, it requires active compliance, with users having to initiate installation, and then upload COVID-19 test results. One of the authors installed a related symptom, called the COVID-19 Symptom Tracker, developed by researchers at Harvard T.H. Chan School of Public Health, Massachusetts General Hospital, King's College London, and Stanford University School of Medicine, in collaboration with Zoe, a health science company. The team notes that "the app asks contributors to answer a few simple questions about themselves and their current health, then to check in every day to report how they're feeling and to list any symptoms they may be having." The app had to be used for several weeks before the app reached enough users (several hundred) in his location (a large US city) to begin reporting with any level of confidence. It is also telling that many institutions are producing multiple apps, initiated by individual groups, with no coherent national-level approach. The US industrial giants Google and Apple Inc. recently released proximity tracing protocol (Exposure Notification API) for their Bluetooth interfaces. Similar to approaches in other countries such as Switzerland, this partnership takes a decentralized approach[66]. Notably, they point out that "if a user chooses to report a positive diagnosis of COVID-19 to their contact tracing app, the user's most recent keys to their Bluetooth beacons will be added to the positive diagnosis list shared by the public health authority so that other users who came in contact with those beacons can be alerted. If a user is notified through their app that they have come into contact with an individual who is positive for COVID-19 then the system will share the day the contact occurred, how long it lasted and the Bluetooth signal strength of that contact." No other information about the contact will be shared and Apple and Google will not receive identifying information about the user, location data, or information about any other devices to which the user has been in proximity.

---

[62] Robeznieks A 2020 Key changes made to telehealth guidelines to boost COVID-19 care Am. Med. Assoc. Online: https://www.ama-assn.org/delivering-care/public-health/key-changes-made-telehealth-guidelines-boost-covid-19-care Retrieved 08 July 2020
[63] Emory School of Medicine 2020 Coronavirus Checker. Online: http://c19check.com/ Retrieved 08 July 2020
[64] Johns Hopkins Medicine 2020. Coronavirus Checker. Online: https://www.hopkinsmedicine.org/coronavirus/covid-19-self-checker.html Retrieved 08 July 2020
[65] Mercy Virtual Care Center FDA 2020 FDA COMBATING COVID-19 WITH MEDICAL DEVICES U.S. Food Drug Adm. Online: https://www.fda.gov/media/136702/download Retrieved 08 July 2020

*16. Brazil:* Brazil's national healthcare system ("Sistema Único de Saúde" or Unified Health System - SUS), is mainly funded by the federal, state and municipal governments through taxes and social contributions. SUS originated after the ratification of the 1988 constitution which granted healthcare as a universal right for all citizens. While public health care is available, it is underfunded and approximately 25% of the wealthier population relies on private insurance companies (Marten *et al* 2014). In April 2020, due to the extraordinary circumstances of COVID-19, the federal government approved law 13.989/2020, which allows the practice of remote consultations with doctors and health care professionals. This measure aimed at reducing the burden on the healthcare system and unnecessary hospital visits. Yet, the approval is for emergency purposes and temporary, and moreover, the government has vetoed the usage of electronic prescriptions over such systems. In parallel, the Brazilian Ministry of Health released a hotline and the COVID-19-SUS app, which enables triage of patients based on their symptoms. The government app contains official statistics on COVID-19 cases, news, information about the disease, its symptoms, and prevention, as well as a feature for searching for nearby health clinics. The app is available for Android and iOS devices and was downloaded over 1 million times from Google Play. Bitmec Health Technologies Limited (Guatemala City, Guatemala) is a technology startup that has developed a digital telemedicine platform that enables cost-effective and scalable primary healthcare services. Through an interactive web portal, and assisted by health workers acting as intermediaries, Bitmec connects patients with remote physicians and is currently undergoing trials of its platform with numerous providers for COVID-19 screening using digital stethoscopes, cameras and pulse oximeters.

*17. Rwanda:* Contact tracing through smartphones is impossible in many parts of the world, particularly in most parts of sub-Saharan Africa, where the average smartphone adoption rate is around 25% and mobile phone subscription rates are 44%[67,68,69]. Symptom checkers through USSD services have potential, but at this point, have not been widely implemented. Remote health monitoring throughout most of Africa is highly limited and consists primarily of remote doctor consultations and education campaigns. In Rwanda has nearly universal healthcare for its thirteen million citizens and implemented a national lockdown after the arrival of the first COVID-19 case. The approach taken by Rwanda has been heralded as a success, involving collaboration between government and the private sector to identify vulnerable groups, set up contact tracing and test 29,395 citizens for COVID-19 by the end of April 2020, where prevalence was 0.7% (Condo *et al* 2020). Rwanda has both a dedicated COVID-19 hotline number (144) and USSD service and the Rwanda National Police have used drones to deliver messages to local communities about how to combat the COVID-19 [70]. A team of state-of-the-art humanoid robots has been deployed to deliver food and medication and screen the temperatures of 50 to 150 people per minute[71]. The Rwanda Utility Regulatory Authority (RURA) has a department focused on the use of mobile data for enabling advanced statistical analysis in collaboration with Carnegie Mellon University Africa (CMU-Africa). Mobile call detail records (CDR) combined with satellite imagery has been used to successfully predict multidimensional poverty (Njuguna and McSharry 2017). The ability to infer the movement of people combined with other sources of information has already been of utility for understanding tourism and trade patterns. Rwanda has deployed a centralized digital contact tracing approach based at RURA[72]. The technology uses mobile phone location tracking enabled by movement analytics to identify people that have come into close contact with confirmed infected cases. Collaboration between CMU-Africa, the Ministry of ICT & Innovation, the National Institute of Statistics Rwanda (NISR), RURA and the Rwanda Biomedical Centre provides potential for using remote screening and machine learning approaches to predict virus hotspots and generate forecasts of the propagation of the virus. There are also plans to develop telemedicine capability to assess suspected cases without the need for these subjects to physically move.

---

[66] Apple 2020 Privacy-Preserving Contact Tracing Apple Online: https://www.apple.com/covid19/contacttracing/ Retrieved 08 July 2020

*18. South Africa:* In South Africa, COVID-19 has led to rapid progress in telemonitoring. In March 2020, the Health Professionals Council of South Africa published updated guidelines lifting restrictions, originally published in 2014, that limited telemonitoring to very specific cases. The new guidelines allow doctors to contact patients, with whom they preferably have an existing relationship, by phone to monitor chronic diseases or follow-up on acute conditions [41]. The private healthcare sector has responded to these new regulations by initiating online COVID-19 consultation services via video chat[42]. Discovery Health, a large health insurance provider, and Vodacom, a major Mobile Network Operator, have partnered to develop a platform that connects

---

[67] Silver L and Johnson C 2018 Majorities in sub-Saharan Africa own mobile phones, but smartphone adoption is modest Pew Res. Cent. Online: https://www.pewresearch.org/global/2018/10/09/majorities-in-sub-saharan-africa-own-mobile-phones-but-smartphone-adoption-is-modest/ Retrieved 08 July 2020

[68] Qelp 2020 Africa smartphone penetration at tipping point Qelp Online: https://www.qelp.com/africa-smartphone-penetration-at-tipping-point/ Retrieved 08 July 2020

[69] Radcliffe D 2018 Mobile in Sub-Saharan Africa: Can world's fastest-growing mobile region keep it up? ZDNet Online: https://www.zdnet.com/article/mobile-in-sub-saharan-africa-can-worlds-fastest-growing-mobile-region-keep-it-up/ Retrieved 08 July 2020

[70] Ashimwe E Rwanda Deploys Drones to Raise COVID-19 Awareness in Communities New Times Online: https://allafrica.com/stories/202004120035.html Retrieved 08 July 2020

[71] AP 2020 Coronavirus - Rwanda: Robots arrive to screen Rwandan patients africanews.com Online: https://www.africanews.com/2020/05/20/coronavirus-rwanda-robots-arrive-to-screen-rwandan-patients/ Retrieved 08 July 2020

[72] Gbenga A M 2020 RWANDA JOINS OTHERS IN USING PHONE DATA FOR COVID-19 CONTACT TRACING Ventur. Africa Online: http://venturesafrica.com/rwanda-joins-others-in-using-phone-data-for-covid-19-contact-tracing/ Retrieved 08 July 2020

patients who have COVID-19-related health questions to a doctor, and have initially made this service free to the public[43]. In order to monitor and reduce the spread of COVID-19, the South African government, through the National Institute for Communicable Diseases, issued regulations that mandated the establishment of a centralized database to facilitate contact tracing (Klaaren *et al* 2020). The centralized contact tracing system termed "COVID-19 Tracing Database" collects and aggregates information from a large range of sources including mobile phone operators, electronic communication services, testing laboratories and accommodation registries (Klaaren *et al* 2020). The information collected in the COVID-19 Tracing Database may only be used by authorized persons and exclusively for the purpose of facilitating contact tracing and monitoring of COVID-19[73]. Contact tracing efforts are further supported by Covi-ID, a mobile QR-code based location tracing application that was developed by the University of Cape Town (UCT) for the South African Government[74]. Covi-ID is a voluntary application that once installed allows users to register themselves and then log their location by scanning a QR-code. All information is stored in a decentralized blockchain and users are in full control of their data. Should a user discover that they are COVID-19 positive, they need to agree to have their health information released before it is accessible to the authorities[75]. COVID-19 has resulted in an increased use of mobile clinics in South Africa. In April 2020, the National Health Laboratory Service in partnership with the National Department of Health rolled out 67 specially fitted vehicles to facilitate COVID-19 swabbing and testing[76]. These vehicles are staffed by two nurses and a driver and act as mobile clinics for COVID-19 testing and contact tracing. In addition to these road vehicles there is also a Railway Clinic called The Phelophepa Train that is now doing COVID-19 testing in remote areas [77,78]. The Phelophepa Train is managed by Transnet (a railway company owned by the South African Department of Enterprises) and supported by funding from several private companies and charities[79,80]. While there is an abundance of COVID-19 information online, it is only available to those who can afford access to data. South Africa has a very high mobile data rate, significantly restricting access to these services[81]. To address this problem, the University of

---

Cape Town (UCT) developed a free USSD-based COVID-19 health literacy education platform and program to provide information to underserved members of society[82].

*19. Australia:* Australia is handling the COVID-19 pandemic in a different way to many parts of the world, especially USA and Europe. The Australian government has enforced social distancing measures much more stringently. Moreover, Australia is very different from these territories with respect to population density, with three persons per square kilometer, as compared to 36 in the USA, and 105 in the European Union [83]. More importantly, one third of Australians live in remote and rural areas. Telemonitoring has been one of the priorities of the Australian government in the pursuit of improved health services for Australians living in rural and remote areas. To support telemonitoring, in 2011, the Australian government introduced an important policy on telemonitoring funding, requiring Medicare to pay specialists for video consultations of remote patients[84]. During the COVID-19 pandemic, telemonitoring has turned into a key weapon in combatting viral spread and in providing continued access to essential primary health services in Australia [85]. The Australian Government expanded the Medicare-subsidized telemonitoring services for all Australians, which was otherwise limited to Australians living in telemonitoring-eligible areas, residents of eligible Residential Aged Care Facilities (RACF) and patients of eligible Aboriginal Medical Services in all areas of Australia. The expansion is now covering services such as general practitioner consultations via telemonitoring, which is limiting unnecessary exposure of patients and health professionals to COVID-19. In addition, this is taking pressure off hospitals and emergency departments, and supporting self-isolation and quarantine policies. Several companies in communication and medical software have supported this volume shift, enabling a seamless transition during this unprecedented time. The National Broadband Network Corporation, which is a Government Business Enterprise, has recently upgraded Australian GP clinics to Internet connectivity speeds of 50/20 Mbps (download/upload) at no extra cost, for six months[86]. This has increased the capability of clinics to deliver multiple consultations through telemonitoring simultaneously. MedicalDirector, a Sydney-based medical software provider, has recently made its fully integrated telemonitoring solution available for free (for at least three months) to support GP clinics. In the first three weeks of use (up to the 20 April 2020), more than 4.3 million consultations were delivered to over three million patients due to the COVID-19 pandemic-driven telehealth expansion[87]. Despite many obvious benefits of telemonitoring, the actual uptake and integration of it into mainstream practice has been slow (Wade *et al* 2014, Caffery *et al* 2018, 2016). This pandemic (COVID-19) has for the first time in Australia provided nationwide access to telemonitoring. For most Australians, this is the first time they are accessing GP services without attending a center. This may have a long-term effect on the uptake of telemonitoring in the post-COVID era, since people can enjoy the same level of services without driving to the clinic, sitting with a set of patients and taking additional leave for a doctor appointment. Besides these telemonitoring initiatives, the Australian Government has launched two mobile applications so far for fighting COVID-19 pandemic. The first mobile application (Coronavirus) was released, at the early stage of COVID-19 in Australia, to provide a single point of contact to check symptoms, register isolation, find advice on different government stimulus packages, access news and media related to COVID-19, identify the current status of COVID-19 across Australia (different states and territories) and essential information (updated regularly). The second mobile application, COVIDSafe, is designed to reduce community exposure from the spread of COVID-19. The app uses the strength of Bluetooth signal for determing the proximity at the mobile phone level (decentralized). The app securely stores (locally) encrypted information of all other app users who are in close proximity. . Besides storing encrypted data into local storage, the app uses a rolling 21-day to maintain data recency. Contacts that occurred outside of the 21-day window are automatically deleted from the user's phone. For the remote and rural Aboriginal and Torres Strait Islander communities, the Australian Government has taken initiative to establish mobile clinics for a rapid coronavirus test. This program brings the testing time down to 45 minutes, which currently takes up to 10 days for this area. The Kirby Institute at University of New South Wales and the International Centre for Point of Care Testing at Flinders University is managing the program. The program is going to use the existing point-of-care testing technology that has been used by the same group with widespread success in rapid testing of sexually transmissible infections in rural communities. A rapid PCR test will be used at the mobile clinics to detect COVID-19 infections in the early phases of illness. In Australia, a Government-centric approach is being used to contain the spread of the virus. Government is at the forefront of all actions and tech industries, universities and all other agencies are playing supporting roles to the government. This may be different from other countries, where many solutions are coming from industry or university alone and then adopted by the government to tackle COVID-19. Telemonitoring played a key role so far in Australia to keep high-quality medical services during this pandemic and the significance of face-to-face consultations for providing better health services is sounding more of a taboo rather than reality. Since the urban population received the most benefit of telemonitoring for the first time, it may encourage integration of telemonitoring services in mainstream healthcare. The experience of all stakeholders (patient, doctors, clinics, government) that will be gained from already provided huge numbers of telemonitoring services are also going to affect the post-COVID-19 telemonitoring structure. In summary, telemonitoring is going to play a big role in shaping healthcare in Australia in the post-COVID-19 era.

---

[82] UCT News 2020. COVID-19 USSD-code health literacy campaign. In UCT News. https://www.news.uct.ac.za/campus/communications/updates/covid-19/-article/2020-04-20-covid-19-ussd-code-health-literacy-campaign Retrieved 08 July 2020

[83] The World Bank Population density (people per sq. km of land area) World Bank Online: https://data.worldbank.org/indicator/EN.POP.DNST Retrieved: No Longer Available

[84] Australian Government Department of Health Telehealth: Specialist video consultations under Medicare Aust. Gov. Dep. Heal. Online: http://www.mbsonline.gov.au/internet/mbsonline/publishing.nsf/Content/mbsonline-telehealth-landing.htm Retrieved: 08 July 2020

[85] Hunt G 2020a Australians embrace telehealth to save lives during COVID-19 Aust. Gov. Dep. Heal. Online: https://www.health.gov.au/ministers/the-hon-greg-hunt-mp/media/australians-embrace-telehealth-to-save-lives-during-covid-19 Retrieved 08 July 2020

[86] Sarcevic A 2020 GP clinics to get free nbn boost for telehealth Govtech Rev. Online: https://www.govtechreview.com.au/content/gov-digital/news/gp-clinics-to-get-free-nbn-boost-for-telehealth-1270288655 Retrieved 08 July 2020

[87] Hunt G 2020a Australians embrace telehealth to save lives during COVID-19 Aust. Gov. Dep. Heal. Online: https://www.health.gov.au/ministers/the-hon-greg-hunt-mp/media/australians-embrace-telehealth-to-save-lives-during-covid-19 Retried 08 July 2020

2*0. New Zealand:* New Zealand is perhaps the most famous success story in containing the pandemic, which like Australia, is geographically isolation and has a low population density. This is not enough to prevent transmission though. The success is largely credited to its government's aggressive lockdown strategy at early stage of the disease spread. The first case of covid-19 in New Zealand was recorded on 28 February and they implemented partial lockdown on 25[th] March with only 205 COVID-19 cases and no deaths. The lockdown was one of the strictest in the world, only permitting people to leave their homes for essential reasons like buying food and going to the doctor. They also have closed the country's borders to non-nationals on 19 March. On June 8, New Zealand lifted all its restrictions except for it border control measures. At the time of writing, the country does not have a single locally acquired case. This is not the only place in which New Zealand has been at the forefront of action in this space though. Orion Health (Wellington, New Zealand) is now offering a free national solution to support scenario modelling, risk prediction, forecasting and planning throughout New Zealand's COVID-19 response. It will not be just the collection of information and point-of-care response that helps reduce disease transmission, but also the use of data science, predictive modelling and the response to those predictions that will help guide appropriate responses. Like many other countries, New Zealand also released a contact tracing app "NZ COVID Tracer" [88]. However, in contrast to other contact tracing app, they create a digital diary of places they visit by scanning QR codes displayed at the entrances to business premises, other organizations and public buildings. The app securely stores the digital diary in the mobile phone and delete automatically aftr 31 days. The information will be pass to the contact tracers only if any person registers their contact information with the National Close Contact Service through the app. At the point of writing, there is no available statistics about how many New Zealander have either downloaded the app or registered through the app. As a part of encouraging telemonitoring during COVID-19, New Zealand government has approved exemption for signature on prescription. However, there is no evidence on its impact on usage of telemonitoring, since during the partial lockdown people were allowed to visit doctors. Usage information of telemonitoring during pre-COVID-19 era is also insufficient to draw any conclusion. One reason behind this could be the successful control of the spread of the disease at the early stages of breakthrough.

**Discussion**

*Remote health monitoring:* Telemedicine does not pretend to replace the classic consultation, but rather it responds to a real need. For example, orientation of patients and prioritizing the degree of urgency of care. This does not mean fewer consultations, but more efficient cooperation, better patient care, and therefore better efficiency of the healthcare system. In addition, remote healthcare services make medicine more accessible to everyone, especially relevant in a context where healthcare is suffering from a sharply declining medical demography and an increase in medical desertification. Overall, telemonitoring has a number of key strengths to enable quality care (e.g. patient inhaler adherence (Taylor *et al* 2018)) and enhanced emergency response. Despite many obvious benefits of telemonitoring, the actual uptake and integration of it into mainstream practice has been slow. In many countries, past talks about remote examinations or consultations with patients were met with suspicion, being considered substandard to in-person visits, if not completely taboo or legally prohibited. France and Germany are good examples of such a situation where the technological as well as some legal and reimbursement frameworks were in place before the pandemic to support teleconsultations. Yet, these were only used marginally. During the pandemic, this has changed dramatically with countries such as France, Italy, Japan and the US enabling new health services to be provided through telemonitoring (Table

---

[88] Anon 2020 NZ COVID Tracer app released to support contact tracing *Minist. Heal.* Online: https://www.health.govt.nz/news-media/media-releases/nz-covid-tracer-app-released-support-contact-tracing

2) and facilitated reimbursement for these services during the time of the pandemic. This has resulted in many patients having now personally experienced telemonitoring. This together with the reported contagion rate of COVID-19 in the clinical environment leading to clusters of nosocomial infections has provided the opportunity for patients to understand the potential risks associated with visiting clinics or hospitals. It is thus expected that COVID-19 pandemic will have a profound impact on the way physicians and healthcare delivery organizations interact with patients and the general population. In particular, face-to-face patient-doctor contacts will be less common and remote monitoring is facilitated by the increasing number of clinical-grade portable medical devices. It is reasonable to imagine the COVID-19 era being a trigger for adopting these long-existing yet never consecrated technologies. However, this statement needs to be moderated by the likely gradual and possibly slow evolution in changing very traditional medical systems in countries such as France and Japan. It also remains unclear how to overcome challenges intrinsic to telemonitoring, e.g., performance of thorough physical assessments and diagnostic tests. Thus, healthcare providers will have to decide what and when patients should be seen remotely versus in person. One other important item that emanated from this review, is that telemedicine tools often lack integrated digital health sensors that may be used to record vital signs (e.g. blood pressure, heart rate). Indeed, although we reviewed many start-ups developing such systems, these are not yet integrated into telemedicine platforms. We made a similar observation for electronic medical record systems which are often not integrated on the telemedicine platforms with the exception of some countries such as Portugal. We thus foresee that an important step forward in telemonitoring will consist of integrating connected biosensors and patients' electronic medical records in secure, dedicated telemedicine platforms. Another interesting observation was that some countries, such as China and Singapore, also created mobile clinics i.e. medical services away from medical centers (Table 3) as another way to provide remote health services in the form of medical services outside the classical clinical environment. Finally, a number of countries such as Israel, the UK and parts of the US, for example, used point-of-contact screening (Table 4) through measurement of symptoms in everyday locations such as at the entry of workplaces (e.g. Figure 2).

It is important to note, that this is not the first time telemonitoring has been used during a pandemic or other extreme situations. For example, China explored the use of telemonitoring in 2003 during the outbreak of Severe Acute Respiratory Syndrome (SARS) (Jia LT, Zhao JQ, Zhang SQ 2020). Teleconsultations were also used in Taiwan for patients with SARS (Keshvardoost *et al* 2020). During the Ebola crisis in 2014-2016, an approach to combat the disease was using a mobile app named Ebola Contact Tracing that supported remote monitoring and contact tracing of confirmed cases of Ebola (Keshvardoost *et al* 2020). In 2017, private telemedicine companies provided services to victims of hurricane Harvey and primary care providers[89]. Given the precedents for the usage of telemonitoring it is important to highlight what are the unique features of the present COVID-19 pandemic that may result in long term adoption of telemedicine tools. In that respect, this review highlighted: (1) the modern legal frameworks that are more inclusive of telemonitoring; (2) modern technological advances for data recording and transmission; (3) the unique global singularity of the pandemic. It is indeed the first time in modern history that the majority of the whole population from so many countries were forced to stay confined at home. This will likely leave a unique psychological mark as well as resulted in many individuals and doctors to experiment with telemedicine services. This is unpreceded.

---

[89] Wicklund E 2020 Harvey's Aftermath Brings mHealth, Telehealth to the Forefront mhealthintelligence Online: https://mhealthintelligence.com/news/harveys-aftermath-brings-mhealth-telehealth-to-the-forefront

*Data protection and privacy concerns:* The introduction of novel technologies in the age of digitalization and big dataraises a number of ethical and privacy concerns, notably in the domain of public health (Vayena and Madoff 2019). Notably, automated face recognition, dystopian tracking of individual's behaviors and identity-theft are major concerns, but the increased potential for being tracked, hacked or *zoom-bombed* during a telemedicine consult, are equally present fears in remote monitoring these days. Even in times of urgent need, important questions on how to deal with personal health data, whether to share this private data for the the public good, and whether this task should be in the responsibility of government or non-government actors are subject to discussion and, at least in democratic countries, needed to be re-negotiated or temporarily installed. Large disparities in privacy-preserving approaches for digital contact tracing can be seen across countries (Table 1). While many early adopters implemented a location tracing with centralized data flows and lower data protection standards, the decentralized approach driven by academia, was adopted in the the Google/Apple Exposure Notification API and gained rapidly more followers, especially in Europe. These interfaces are the only ones that enable effective decentralized Bluetooth proximity tracing on the respective smartphone devices and this dependency lead for example Germany and the UK (also Austria, not discussed in this manuscript) changing tack to switch from a centralized to a decentralized approach. It remains to be seen what effect the industry driven decentralized proximity tracing will have on public perception in various countries. Resistance against the smartphone OS duopoly has clearly manifested in France where, as a consequence, the centralized contact tracing app is not compatible with solutions of neighboring countries such as Germany and Switzerland. Despite the higher privacy protection standards, many countries required legal changes and parliament approval, before introduction to the public, partially explaining why countries with high privacy expectation had a relatively late introduction of digital contact tracing solutions. Because of these political and technical challenges, decentralized native phone OS software is probably the best solution for implementing a robust infrastructure for digital contact tracing. With the absence of government leadership in this domain in many countries, as seen in the US with many incompatible approaches circulating in various states, an industry dominated decentralized proximity tracing may be the only hope of providing transnational, privacy-preserving large-scale contact tracing, but also illustrates the dependence on these industrial players.

All organizations that are collecting, processing and storing medical data must comply with international and local regulations. For the USA, the data protection law that is required is the Health Insurance Portability and Accountability Act (HIPAA). The European Union's General Data Protection Regulation (GDPR) is a regulation in the European Union (EU) on data protection and privacy within the EU. It also regulates aspect of personal data exportation outside the European Economic Area (EEA). These regulations have important implications within the context of medical data processing. GDPR specifies actions to be taken such as: data holders must notify the authorities in case of breach, subjects have a right to access their data, the right to be forgotten that gives the individuals the possibility to request the removal of their personal data, privacy must be designed in and included by default (Vitabile *et al* 2019). In the case of noncompliance severe fine may be imposed. Within the scope of the pandemic there has been debates about how GDPR will influence data sharing and analysis of digital health data collected in Europe for research purposes (McLennan *et al* 2020). This includes health data collected remotely. It has been put forward that the research exemption clause of the GDPR during COVID-19 should be used in order to facilitate research efforts using these data. This highlights important considerations relating to data protection laws and some of their implications within the area of remote health monitoring and how these laws may have exemption clauses that enable special usage of data under GDPR in certain extraordinary circumstances. Whether this will include the current pandemic is yet to be determined but we agree with the general call (McLennan *et al* 2020) for making these regulations more flexible within the context of the COVID-19 pandemic as far as it may benefit research and human health while preserving privacy.

*Main innitiators*: Most of the countries had pandemic handling strategies in place before COVID-19 pandemic. As a result, government is as expected the main initiator in the surveillance and tackling of any pandemic. However, use of the state-of-art technologies and technological innovation to help or improve the process brings the opportunity for industry and academia to contribute on top of the government agencies. In most of the studied regions, government is the initiator (either sole or with the industry) to introduce contact tracing app except for Switzerland and the United States (Table 1). In Switzerland, the app is introduced by the academia and in United States it is both by academia and the industry. Irrespective of the region, users are concerned about privacy and data protection features of contact tracing apps, which hindered the wide acceptance of such apps. At the time of writing this part, no solid evidence is reported on how these apps have helped so far in tracking or tackling the pandemic. Usage of telemonitoring mostly increased across all studied regions (Table 2). It is interesting that in only a few regions, including Japan, Italy, Russia, Canada, Brazil and Australia, the government is the main initiator to encourage usage of telemonitoring. For regions where industry is the main initiator, it can be assumed that the penetration of telemonitoring in non-pandemic conditions is also higher, which resulted in a stable business model, where industry can play a significant role. Reduced industry participations in some countries can also be contributed to strong public health systems or government centric health care systems. An interesting distribution in initiating mobile clinics is seen in the studied regions (Table 3). Approximately one third of the regions have not introduced mobile clinics, one third were initiated by the government and the other one third is by the industry. Most of the European countries are found to be the one which had not introduced any mobile clinic, it may be attributed to the already existing adequate healthcare facilities with respect to demand. Another reason for not having any mobile clinics can be attributed to the lack of reserve resources to run such additional facilities on top of existing ones. It is also interesting to see that industry initiated mobile clinics are mostly in Asia and Africa The point-of-contact screening for COVID-19 symptom is initiated by the government in half of the studied region. Among the rest, in four regions it was initiated by the industry sector. Only Singapore among the industry initiated point-of-care screening regions has shown high usage, otherwise mostly low. In contrast, in government initiated regions, most of them have shown high usage except South Africa, Canada and France.

*Lack of evidence:* Perhaps the most pressing issue in remote monitoring in this time of crisis, is the almost total lack of evidence of the benefit for much of the existing and newly developed technology that were used during the pandemic. Much of it is promising, but by definition, it has largely remained untested in the current paradigm. With companies driving the technology (because they are most prepared to scale), secrecy and the need to protect proprietary information often impedes knowledge discovery, and marketing hype replaces hard science. Regulatory bodies have been fast-tracking COVID-19-relevant applications, in the hope of addressing this issue. Outside the purview of the regularity bodies, repurposed technology presents different issues, particularly through the heterogeneity of sensors and software. Bluetooth technology for contact tracing is a case in point. Currently, there is very little evidence to support the guidelines for the exposure estimate of less than 2 meters for 15 minutes as a trigger for initiating a contact trace. For instance, Google and Apple have left it to the manager of the system to decide based on state-wide policies in the US, although they have set a default, which most people will likely use. Perhaps more worryingly, there is little evidence to suggest that we can accurately estimate the distance between two Bluetooth sensors in an unknown environment, where ambient radiofrequency conditions, and physical objects such as walls, could create substantial numbers of false positives and false negatives. It is too early to judge whether or not contact tracing apps were effective, in particular because the launch of these apps also coincided with lockdowns and other special policies taken to counter the virus. It is also challenging to estimate how telemedicine approaches compares to conventional care in the COVID-19 era, given that no such telemedicine alternatives were used at such a scale beforehand. However, it was sometimes the only alternative due to the perceived fear of hospital infection discouraging patients from going to the hospital even though they suffered from severe illnesses. Yet, outside of the COVID-19 era telemedicine approaches had already shown some successes when compared to their conventional counterpart such as for heart failure management in reducing all cause hospitalization and a number of other clinical endpoints (Zhu et al 2020) and stroke management (Guzik and Switzer 2020).

*Conclusions:* In summary, the need for testing, symptom triage, contact tracing, remote monitoring of mild COVID-19 patients, together with targeted movement restriction at early stages is driving most of the innovation in remote monitoring in the time of a pandemic. It is difficult to innovate beyond this paradigm without creating extra burden or running head-long into privacy issues that may ultimately defeat the original positive intentions. Moreover, the quality, accuracy and utility of heart rate, respiration rate and oxygen saturation levels from video cameras has yet to be proven but certainly has real potential, with many telemedicine providers seriously considering this technology as an adjunct to video review. More advanced techniques, such as detection of pain from facial imagery, and abnormal behaviors from movement via video or cellphone usage may provide utility in the future as additional covariates to increase specificity. However, despite the best efforts of entities like Babylon Health and the Human Diagnosis Project, we are still a long way from the automated "artificially intelligent" consult or even a COVID-19 specific screening and triage system based on standard machine learning. Indeed, relatively simple self-report symptom checkers may be one of the most useful remote health technologies. Never-the-less, this current pandemic appears to be ushering in a new era for remote monitoring, which will take months or even years to play out against a landscape of shifting perceptions, regulatory pressures, financial constraints, and medical needs.

**Table 1:** Contact tracing including proximinty and location tracing across countries.

| Country | Increased usage of remote health monitoring | Main initiator | Personal Data Protection |
|---|---|---|---|
| China (1) | High | Government | Low |
| Singapore (2) | High | Government | Moderate |
| India (3) | High | Government | Low |
| South Korea (4) | High | Government | Low |
| Japan (5) | High | Government | Strong |
| Israel (6) | High | Government | Low |
| United Kingdom (7) | High | Government & Industry | Moderate |
| France (8) | High | Government & Industry | Strong |
| Italy (9) | High | Government & Industry | Strong |
| Germany (10) | High | Government | Strong |
| Switzerland (11) | High | Academia | Strong |
| Portugal (12) | n/a | n/a | n/a |
| Russia (13) | No effect | n/a | n/a |
| Canada (14) | High | Government & Industry | Moderate |
| United States (15) | High | Industry & Academia | Strong |
| Brazil (16) | No effect | n/a | n/a |
| Rwanda (17) | High | Government | Low |
| South Africa (18) | High | Government | Low |
| Australia (19) | High | Government | Moderate |
| New Zealand (20) | No effect | n/a | n/a |

**Table 2:** Telemonitoring (usage of telemedicine tools) across countries.

| Country | Increased usage of remote health monitoring | Main initiator | Personal Data Protection |
|---|---|---|---|
| China (1) | High | Industry | Moderate |
| Singapore (2) | Moderate | Industry | Strong |
| India (3) | Moderate | Industry | Moderate |
| South Korea (4) | High | Industry | Moderate |
| Japan (5) | High | Government | Strong |
| Israel (6) | High | Industry | Moderate |
| United Kingdom (7) | High | Industry | Moderate |
| France (8) | High | Industry | Strong |
| Italy (9) | High | Government | Moderate |
| Germany (10) | High | Industry | Strong |
| Switzerland (11) | Moderate | Industry | Moderate |
| Portugal (12) | Moderate | Government & industry | n/a |
| Russia (13) | High | Government | Unknown |
| Canada (14) | High | Government | Strong |
| United States (15) | High | Industry | Unknown |
| Brazil (16) | High | Government | Unknown |
| Rwanda (17) | n/a | n/a | n/a |
| South Africa (18) | High | Government & Industry | Moderate |
| Australia (19) | High | Government | Strong |
| New Zealand (20) | n/a | n/a | n/a |

**Table 3:** Creation of mobile clinics i.e. medical services away from medical centers.

| Country | Increased usage of remote health monitoring | Main initiator |
|---|---|---|
| China (1) | High | Industry |
| Singapore (2) | High | Industry |
| India (3) | Low | Industry |
| South Korea (4) | Low | Industry |
| Japan (5) | High | Government |
| Israel (6) | Low | Industry |
| United Kingdom (7) | No effect | n/a |
| France (8) | No effect | n/a |
| Italy (9) | Low | Government |
| Germany (10) | No effect | n/a |
| Switzerland (11) | No effect | n/a |
| Portugal (12) | n/a | n/a |
| Russia (13) | High | Government |
| Canada (14) | Low | Government |
| United States (15) | Low | Industry |
| Brazil (16) | n/a | n/a |
| Rwanda (17) | High | Government |
| South Africa (18) | High | Government |
| Australia (19) | High | Government & Industry |
| New Zealand (20) | No effect | n/a |

**Table 4:** Point-of-contact screening through measurement of symptoms in everyday locations.

| Country | Usage of remote health monitoring | Main initiator |
|---|---|---|
| China (1) | High | Government |
| Singapore (2) | High | Industry |
| India (3) | High | Government |
| South Korea (4) | Low | Industry |
| Japan (5) | Moderate | Government |
| Israel (6) | High | Government |
| United Kingdom (7) | High | Government |
| France (8) | Low | Government |
| Italy (9) | High | Government |
| Germany (10) | No effect | n/a |
| Switzerland (11) | No effect | n/a |
| Portugal (12) | n/a | n/a |
| Russia (13) | Moderate | Industry |
| Canada (14) | Low | Government |
| United States (15) | Low | Industry |
| Brazil (16) | High | Government |
| Rwanda (17) | n/a | n/a |
| South Africa (18) | Low | Government |
| Australia (19) | Low | Industry |
| New Zealand (20) | No effect | n/a |

**Conflicts of interest**

GC has financial interest in Alivecor Inc and receives unrestricted funding from the company. GC also is the CTO of Mindchild Medical and has ownership interests in Mindchild Medical. WK declares no conflict of interest. WK and Leitwert GmbH collaborate in an unrelated research project. MO is an employee of Babylon Health. FA is employed by Sensyne Health. PAW is an employee of PeriGen, a developer and provider of decision-support systems for obstetrical car